\documentclass{article}

\usepackage[preprint]{neurips_2024}

\usepackage[utf8]{inputenc} % allow utf-8 input
\usepackage[T1]{fontenc}    % use 8-bit T1 fonts
\usepackage{hyperref}       % hyperlinks
\usepackage{url}            % simple URL typesetting
\usepackage{booktabs}       % professional-quality tables
\usepackage{amsfonts}       % blackboard math symbols
\usepackage{nicefrac}       % compact symbols for 1/2, etc.
\usepackage{microtype}      % microtypography
\usepackage{xcolor}         % colors

\usepackage{graphicx}
\usepackage{multirow}
\usepackage{bm}

\usepackage{comment}
\usepackage{algorithm}
\usepackage{algorithmicx}
\usepackage{algpseudocode}
\usepackage{color}
\usepackage{amsmath}
\usepackage{autobreak}
\usepackage{amssymb}
\usepackage[capitalize,noabbrev]{cleveref}

\usepackage{graphicx}
\usepackage{subcaption}

\usepackage{pifont}       % \ding{xx}
\usepackage{bbding}       % \Checkmark,\XSolid,... (需要和pifont宏包共同使用)
\usepackage{fontawesome}  % \faCheck,\faTimes
 
\newtheorem{proposition}{Proposition}
\newtheorem{propositionn}{Proposition}

\title{
Regressor-free Molecule Generation \\ to Support Drug Response Prediction
}

\author{
  Kun Li \\
  School of Computer Science\\
  Wuhan University\\
  \texttt{li\_\_kun@whu.edu.cn} \\
  \And
  Xiuwen Gong \\
  University of Technology Sydney \\
  \texttt{gongxiuwen@gmail.com} \\
  \AND
  Shirui Pan \\
  School of Information and Communication Technology \\
  Griffith University \\
  \texttt{s.pan@griffith.edu.au} \\
  \And
  Jia Wu \\
  School of Computing \\
  Macquarie University \\
  \texttt{Jia.wu@mq.edu.au} \\
  \And
  Bo Du \\
  School of Computer Science\\
  Wuhan University \\
  \texttt{gunspace@163.com} \\
  \And
  Wenbin Hu \thanks{Corresponding author} \\
  School of Computer Science\\
  Wuhan University \\
  \texttt{hwb@whu.edu.cn} \\
}

\begin{document}

\maketitle

\begin{abstract}

Drug response prediction (DRP) is a crucial phase in drug discovery, and the most important metric for its evaluation is the IC50 score. DRP results are heavily dependent on the quality of the generated molecules. Existing molecule generation methods typically employ classifier-based guidance, enabling sampling within the IC50 classification range. However, these methods fail to ensure the sampling space range's effectiveness, generating numerous ineffective molecules. Through experimental and theoretical study, we hypothesize that conditional generation based on the target IC50 score can obtain a more effective sampling space. As a result, we introduce regressor-free guidance molecule generation to ensure sampling within a more effective space and support DRP. Regressor-free guidance combines a diffusion model's score estimation with a regression controller model's gradient based on number labels. To effectively map regression labels between drugs and cell lines, we design a common-sense numerical knowledge graph that constrains the order of text representations.  Experimental results on the real-world dataset for the DRP task demonstrate our method's effectiveness in drug discovery. The code is available at: \url{https://anonymous.4open.science/r/RMCD-DBD1}.

% And can facilitate a wider range of applications of condition-based diffusion models for programmed control of molecular maps. 
% We test the drug-cell response conditional controls for stable diffusion.
% RTDG combines molecular diffusion and tag-drug contrast learning models, then fine-tuned for specific tasks. 

% Moreover, training only on small task-specific datasets can greatly affect sampling effectiveness. For this reason, we propose a dual-branch controlled noise prediction method to ensure the effectiveness of the diffusion process and sample. The method connects the network structure of the original diffusion model with zero convolutions, which learns various conditional controls by gradually adding zero-initialized convolutional layers to ensure effectiveness in drug-specific tasks.

% Additionally, we introduce a dual-branch controlled noise prediction method, where the two branches conduct unconditional training and conditional mixed training respectively, to enhance noise prediction performance.

\end{abstract}

\section{Introduction}

% The challenge of the drug discovery process stems from the large and discrete search space for chemical molecules \citep{drug-maker}. Specifically, the possible structural scale of drug-like compounds ranges from $10^{23}$ to $10^{60}$, but a small percentage of these (about $10^{8}$) are therapeutically relevant \citep{drug-like-chemical-space-1,drug-like-chemical-space-2}.

Drug response prediction (DRP) is crucial in drug discovery \citep{drp}. It evaluates drug response in cell lines to aid the screening potential pharmacologically active compounds. However, the challenge in drug discovery arises from the large and discrete chemical molecule search space \citep{drug-maker}. Specifically, the possible structural scale of drug-like compounds ranges from $10^{23}$ to $10^{60}$, but a small percentage of these (about $10^{8}$) are therapeutically relevant \citep{drug-like-chemical-space-1,drug-like-chemical-space-2}. Traditional drug discovery techniques typically involve screening large molecule libraries, resulting in a low probability of finding molecules with an adequate efficacy for specific cell lines.

Moreover. molecule quality directly impacts drug screening efficiency and progress during DRP tasks. Typically, high-quality molecules exhibit enhanced pharmacological properties and are more likely to become effective drug candidates. Therefore, generating high-quality molecules has become a core issue in current drug discovery research. With the development of artificial intelligence technology, various molecule generation methods have been proposed, including sequence-based generative models \citep{seq_based}, variational auto-encoders \citep{vae1,vae2}, normalizing flows \citep{flow1,flow2}, and diffusion \citep{CDGS,GruM2D,E3Diffusion}. Existing molecule generation methods can generate molecules with specific stability and novelty, providing a larger library of virtual screening molecules for drug discovery. 

% The traditional drug discovery process can be divided into two stages: one is to create a library of molecules to be screened, and the other is to predict the properties of molecules virtually. Finally, wet experiments are performed on the virtually screened molecules to verify the properties of the molecules. For task-specific molecular screening, the traditional approach is to screen from existing libraries of drug compounds by the specific model. However, the variety and number of compounds in existing drug compound libraries are relatively small, and screening without a library of molecules designed for a specific task is like looking for a needle in a haystack, which is both time-consuming and inefficient. In the long run, the ideal approach to drug discovery should be de novo molecular design with special conditions.

\begin{figure}
    \centering
    \includegraphics[width=1.0\linewidth]{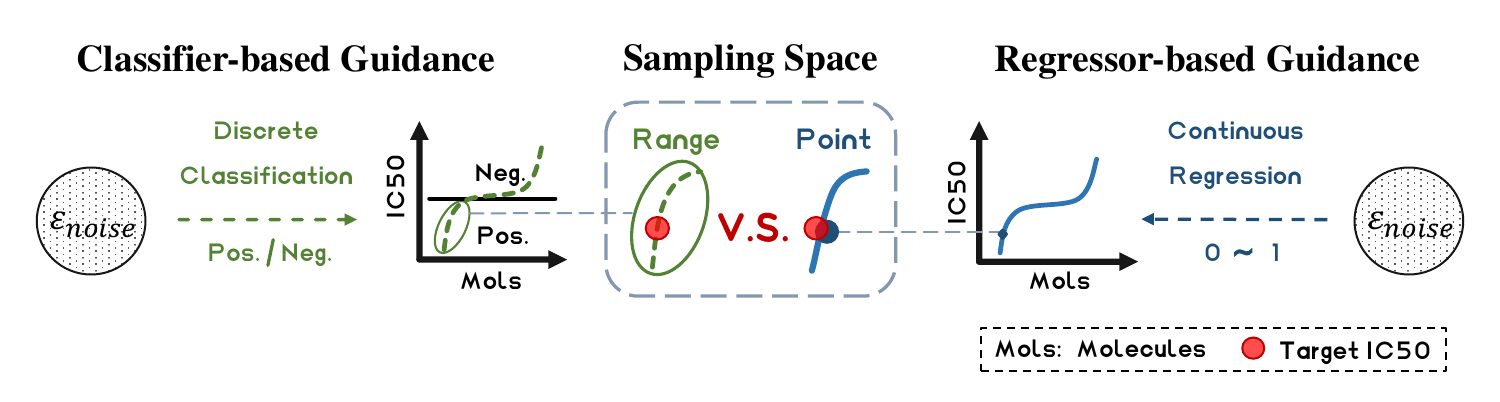}
    \caption{Sampling space comparison for target conditions in classifier- vs. regressor-based guidance molecule generation. }
    \label{fig:1}
\end{figure}

To meet the DRP-based screening tasks' demands, diffusion-based conditional generation methods have garnered significant attention due to ability to generate specific molecules while maintaining result diversity and novelty. As shown in Fig. \ref{fig:1}, traditional diffusion models can generate molecules under certain conditions \citep{CDGS, MOOD, FREED}, which are mostly classifier-based \citep{cls_protein_guided} (detailed related work is referenced in Appendix \ref{B}). However, drugs' features are mostly continuous and quantifiable \citep{cq1, cq2}. The sampling space range guided by the classifier (i.e., whether there is a reaction between the drug and the cell line) is relatively large, hindering its ability to perform precise molecular feature space sampling. In contrast, the diffusion model guided by regressor-based conditions can obtain samples within a narrower range near specific response values. This ensures that the generated molecules are consistent with the target response values, instead of falling within a broader response range. As a result, when generating an identical number of molecules, the number of molecules meeting a given criteria generated using regressor-based guidance is higher than those generated with classifier-based guidance. This improves generation efficiency and increases the drug discovery success rate.

To achieve this, we propose the \textit{regressor-free guided molecule generation} method to ensure sampling within a more effective space, supporting the DRP task. To avoid potential gradient-based adversarial attacks, we opt for a regressor-free guidance method, inspired by classifier-free guidance \citep{Classifier-Free}. Regressor-free guidance is a score-based diffusion scheme, that incorporates a regression controller model, which is based on the number label, into a generative stochastic differential equation (SDE) with a conditional hyperparameter. To effectively map the labels between the drug and the cell line, we create a regression controller model that converts the drug information labels into text with a common-sense numerical knowledge graph (CN-KG). The CN-KG restricts the text representation order. Moreover, since the DRP dataset contains only a limited variety of molecules, training solely on this small dataset can significantly impact the sampling effectiveness. Therefore, we propose a dual-branch controlled noise prediction (DBControl) model to ensure the diffusion processing and sampling effectiveness. The control model is designed for score estimation and consists of two identical graph neural networks (GNNs) connected by zero convolution layers. These convolution layers are gradually added to learn various conditional controls, ensuring the DRP task's effectiveness. 

The experimental results on the GDSCv2 \citep{GDSC} dataset for the DRP task demonstrate out method's effectiveness in de novo drug design.
\textbf{This paper's main contributions are summarized as follows:}
% Regressor Guidance Diffusion for molecule generation
   % \item We propose a natural language-based drug molecular contrast learning as a conditional diffusion model fine-tuning method, which is the first molecular diffusion model for drug-x type multitask conditions.
    % \item We use a zero-convolution layer stepwise control fine-tuning strategy to fine-tune a diffusion model trained on an unconditional large molecular data set.
    % \item A regressor guidance molecular diffusion is proposed for molecular conditional generation, where the regression controller can detect the numerical relationship between a molecule and another object as a conditional guidance for molecule generation.
    % \item To address the over-fitting and poor sampling effect due to the small amount of data in a specific domain, a dual-branch controlled noise prediction model is proposed to control the mixed diffusion of unconditional and conditional datasets through zero-convolution layers.
    % \item Experimental results demonstrate that our method outperforms the state-of-the-art baselines in conditional molecular graph generation based on regression relationships. In addition, we provide a detailed demonstration of the effectiveness of this method.

    % Regressor-free guidance combines the score estimate of a diffusion model with the gradient of a regression controller based on number labels.

\begin{itemize}
    \item Regressor-free guidance molecule generation is proposed to ensure sampling within a more effective space. Regressor-free guidance is a score-based diffusion scheme, that incorporates a regression controller model, which is based on the number label, into a generative SDE with a conditional hyperparameter.
    \item To enhance noise prediction performance, we introduce the DBControl model for score estimation. The control model consists of two GNN-based branches, each undergoing unconditional and conditional mixed training, respectively.
    \item The experimental results demonstrate that our method outperforms state-of-the-art baselines in conditional molecular graph generation for the DRP task. Furthermore, we validate our method's effectiveness.
\end{itemize}

% 一个Regressor Guidance Diffusion 被提出for molecule generation，其中Regression controller可以感知分子与另外一个对象之间的数值关系，作为条件引导分子生成。

% 为解决特定领域数据量小导致的过拟合和采样效果较差，一个双分支控制的噪声预测模型被提出，通过zero-convolution layers控制无条件数据集和有条件数据集混合扩散。

% Experimental results demonstrate that our method outperforms the state-of-the-art baselines in 基于回归关系的条件 molecular graph 生成. 此外，我们为此方法的有效性提供了证明。

\section{Methods}

\begin{figure*}[!ht]
    \centering
    \includegraphics[width=0.95\linewidth]{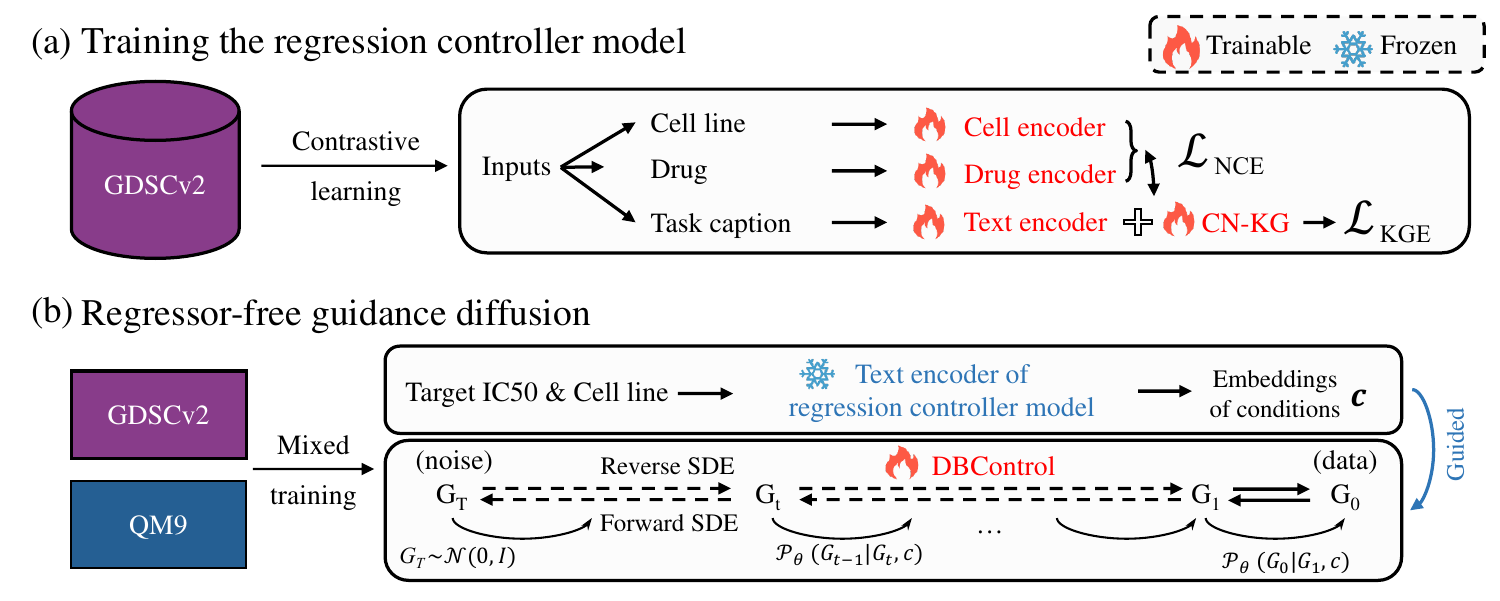}
    \caption{(a) illustrates the training process of the regression controller model, which serves as a conditional encoder guiding diffusion. (b) depicts the regressor-free guidance diffusion process, utilizing the text encoder of the trained regression controller model to encode the target conditions. The DBControl model is a score-based noise prediction model trained on a mixture of the conditional GDSCv2 and unconditional QM9 dataset.}
    \label{fig:framework_our_method}
\end{figure*}

\subsection{Notations} 

\noindent \textbf{Molecular graph representation.} The molecule can be represented as a graph 
$\bm{G}  = (\bm{X},\bm{A} )\in \mathbb{R} ^{N\times F}\times \mathbb{R} ^{N\times N}  \colon = \mathcal{G} $, where $\bm{X}$ and $\bm{A}$ are the node features and weighted adjacency matrix of the graph $\bm{G}$ with $N$ nodes. $\bm{X}$ is the node feature matrix for atom types described by $F$-dimensional one-hot encoding.

% \noindent \textbf{Regression Perception and Relationxx-aware}  

\noindent \textbf{Score-based graph generation.} Based on the seminal work of \citep{songyang}, Graph Diffusion via the System of SDEs (GDSS) was introduced \citep{GDSS}. GDSS utilizes two SDE models to simulate node feature and adjacency matrix diffusion, thereby capturing the complex dependencies between nodes and edges. Thus, the forward diffusion for a graph $\bm{G}$ is defined by an It\^{o} SDE:

\begin{equation}
\label{eq1}
\mathrm{d}\bm{G}_t = \mathbf{f}_t(\bm{G}_t)\mathrm{d}t + \mathbf{g}_t(\bm{G}_t)\mathrm{d}\mathbf{w}, \;\;\; \bm{G}_0\sim p_{data}
\end{equation}

\noindent where $\mathbf{f}_t(\cdot): \mathcal{G}\to \mathcal{G}$ is the linear drift coefficient, $\mathbf{g}_t(\cdot): \mathcal{G}\to  \mathcal{G} \times \mathcal{G}$ is the diffusion coefficient, and $\mathbf{w}$ is the standard Wiener process. The $t$-subscript denotes a time function: $F_t(\cdot):= F(·, t) $, which $t \in [0, T] $. If the initial state distribution is supposed as $p_{0}(\bm{G}_0|\mathbf{c})$, the density at time $t$ is $p_{t}(\bm{G}_t|\mathbf{c})$ when conditioned on $\mathbf{c}$. Therefore, the reverse-time SDE from $T$ to 0 corresponding to Eq.~\ref{eq1} is given by \citep{gradfather}:
\begin{equation}
\label{eq2}
\mathrm{d}\bm{G}_t =  \left \{ \mathrm{\mathbf{f} }_t(\bm{G}_t)-g_t^2\left [\nabla_{\bm{G}}\mathrm{log}\;p_{t}(\bm{G})+  \nabla_{\bm{G}}\mathrm{log}\;p_{t}(\mathbf{c} |\bm{G}_t) \right ]  \right \}  \mathrm{d}\bar{t}+g_t\mathrm{d}\mathbf{\bar{w}},
\end{equation}
\noindent where $\nabla_{\bm{G}}\mathrm{log}\;p_{t}(\bm{G}_t)$ is the graph score function and $\mathbf{\bar{w}}_t$ represents the reverse-time standard Wiener process. The marginal distribution under forward diffusion is denoted as $p_t$, and the corresponding reverse diffusion process can be described by the following SDEs, which describe the diffusion process of each component, $\bm{X}$ and $\bm{A}$, respectively:

\begin{equation}
\label{eq:reverse-time-SDEs}
\left\{ 
    \begin{array}{lc}
        
         \mathrm{d}\bm{X}_t = \left [ \mathbf{f}_{1,t}(\bm{X}_t)-g_{1,t}^2  \nabla_{\bm{X}_t} \mathrm{log}\; p_t(\bm{X}_t, \bm{A}_t) \right ]  \mathrm{d}\bar{t} + g_{1,t}\mathrm{d}\bar{\mathbf{w}}_{1}\\ [5pt]
         \mathrm{d}\bm{A}_t = \left [ \mathbf{f}_{2,t}(\bm{A}_t)-g_{2,t}^2  \nabla_{\bm{A}_t} \mathrm{log}\; p_t(\bm{X}_t, \bm{A}_t) \right ]  \mathrm{d}\bar{t} + g_{2,t}\mathrm{d}\bar{\mathbf{w}}_{2}
    \end{array}
\right.
\end{equation}

\noindent where $\mathbf{f}_{1,t}$ and $\mathbf{f}_{2,t}$ are linear drift coefficients satisfying $\mathbf{f}_t(\bm{X},\bm{A}) = \left(\mathbf{f}_{1,t}(\bm{X}), \mathbf{f}_{2,t}(\bm{A}) \right)$, $g_{1,t}$ and $g_{2,t}$ are scalar diffusion coefficients, and $\bar{\mathbf{w}}_{1}$, $\bar{\mathbf{w}}_{2}$ are reverse-time standard Wiener processes.

% The seminal work \citep{SDE} models the diffusion from data to noise using the stochastic differential equation (SDE). However, their simplistic approach to graph generation fails to capture the complex dependencies between nodes and edges, which are crucial for learning the distribution of graphs. To address this issue, \citep{GDSS} proposes Graph Diffusion via the System of SDEs (GDSS), employing two SDE models to respectively model the diffusion of node features and the adjacency matrix.

\subsection{Method overview}

In this paper, we introduced a regressor-free guidance molecule generation method to ensure sampling within a more effective space, which can support the DRP task. Our method mainly consists of two models: the regression controller model for guiding conditional generation and the DBControl model for score estimation. Our method's framework is shown in Fig. \ref{fig:framework_our_method}. In the diffusion phase, the DBControl model was used for noise prediction. First, DBControl is trained using molecular datasets such as QM9 \citep{QM9} and ZINC250k \citep{ZINC250k}. Then, the regression controller model was trained on the GDSCv2 dataset for the DRP task (i.e., approximately 1\text{\textperthousand} of the molecule number in the unconditional molecule dataset). Finally, the DBControl model was trained on a mixture of the GDSCv2 and unconditional molecular datasets.

\subsubsection{The regression controller model}
\label{sec:regcontroller}

With the regression controller model, a natural language process description on the molecule with the cell line is taken as the text input, and molecule and cell line regular representation (e.g., sequence and motifs encoding) is used as the regular input. Then, the natural language description is generated using templates as follows:

\begin{equation}
    \mathrm{The} \;   \mathrm{response} \;\mathrm{value}\; \mathrm{of} \;\mathrm{the}\; \mathrm{drug} \;\mathrm{with}\; \mathrm{the}\; \mathrm{ {\color[RGB]{63, 114, 175}{\left [ \mathbf{name\; of\;the\;cell\;line}  \right ] }}} \;\mathrm{is} \;\mathrm{ {\color[RGB]{63, 114, 175}{\left [ \mathrm{IC}_{50}  \right ] }}},
\end{equation}

\noindent where $\mathrm{ {\color[RGB]{63, 114, 175}{\left [ \mathbf{name\; of\;the\;cell\;line}  \right ] }}}$ refers to the name of the specific cell line, and $\mathrm{ {\color[RGB]{63, 114, 175}{\left [ \mathrm{IC}_{50}  \right ] }}}$ represents the specific $\mathrm{IC}_{50}$ score. $\mathrm{IC}_{50}$ stands for Half Maximal Inhibitory Concentration, which is commonly used to measure the biological activity of a drug. In this instance, the $\mathrm{IC}_{50}$ score is a concrete number, such as, \textit{"zero five two one"} means that the $\mathrm{IC}_{50}$ score is \textit{0.521}. Furthermore, we used text to describe the value, instead of a string of numbers, such as "0.521". The generated numeric text retains a fixed precision (e.g., 0.521 has a precision of $\xi=3$); the number of valid digits is $\xi \in \mathbb{N}$.

% 药物和X的常规表征方式有很多，例如MLP，CNN，GNN，Transformer等，本文所提出的Regression controller对此不作限制。用$\Phi_{\mathrm{f}$表示常规表征分支， 用$\Phi_{\mathrm{cap}}$表示文本描述分支。

Various drug and cell line representation methods exist, such as  GNNs \citep{GruM2D} and the transformer \citep{DeepTTA}. The regression controller model proposed in this paper has no restriction in this regard; hence, it is not specified. In addition, the regular representation branch is denoted by $\Phi_{\mathrm{f}}$ and the transformer-based textual description branch is $\Phi_{\mathrm{cap}}$. The $\Phi_{\mathrm{cap}}$ input is the target $\mathrm{IC}_{50}$ scores and cell lines, denoted as $c=\left ( \mathbf{X_{cell}},\mathbf{X_{number}} \right ) $. We normalized the feature vectors in a hyper-sphere using $u_i:=\frac{ \Phi_{\mathrm{f}}(\mathbf{Drug}, \mathbf{X} )}{\left \|  \Phi_{\mathrm{f}}(\mathbf{Drug},\mathbf{X} ) \right \| }$ and $v _j:=\frac{ \Phi_{\mathrm{cap}}(\mathbf{X_{cell}}, \mathbf{X_{number}} )}{\left \|  \Phi_{\mathrm{cap}}(\mathbf{X_{cell}}, \mathbf{X_{number}} ) \right \| } $. Then, the similarity between $u _i$ and $v _j $ was calculated as $u_i^\mathrm{T} v _j$. Finally, the contrast learning between the two branches was optimized by the supervised contrastive loss function. 

\begin{equation}
    \label{eq:nce}
    \begin{aligned}
        \mathcal{L}_{\mathrm{NCE}} = -\frac{1}{N} \left ( \sum_{i}^{N}\mathrm{log} \frac{\mathrm{exp} ( u_i^\mathrm{T} v _i/\sigma )}{  {\textstyle \sum_{j=1}^{N}} \mathrm{exp} ( u_i^\mathrm{T} v _j/\sigma )} + \sum_{i}^{N}\mathrm{log} \frac{\mathrm{exp} ( v_i^\mathrm{T} u_i/\sigma )}{  {\textstyle \sum_{j=1}^{N}} \mathrm{exp} (v_i^\mathrm{T} u_j /\sigma )} \right ),
    \end{aligned}
\end{equation}

\noindent where, $N$ is the batch size, and $\sigma$ is the temperature value used for scaling the logits. 

% These CN-KG entities are integers, the significant number of $\mathbf{c}$ multiplied by the specified precision, where the minimum and maximum numbers are customized for the specific task.

Therefore, to enable the text module to represent the numerical labels at the standard ordered cognition level, we introduced the CN-KG \citep{CLDR} to enhance the ordered representation of numerical texts. The CN-KG has natural number entities, denoted as $E \subseteq \mathbb{N}^{*} $. The $E$ entities are linked by a single relationship type called "is less than," denoted as $L$, which ensures that number transfer properties are captured.  Furthermore, the margin-based loss function $\mathcal{L}_{\mathrm{KGE} }$ was used for the CN-KG embedding. We aimed to minimize the differences in embedding vectors between the entities' set $E$ and the single relationship $L$ (i.e. is less than) 

\begin{equation}
\label{eq:kge}
\mathcal{L}_{\mathrm{KGE}}= \sum_{ (h,l,t)\in S   } \left [ \gamma +d(\textbf{h}+\textbf{l},\textbf{t}) -d(\textbf{t}+\textbf{l} ,\textbf{h})\right ]_+,
\end{equation}

\noindent where, $[x]_+$ denotes the positive part of $x$, $\gamma > 0$, and is a margin hyperparameter. The set $S$ is composed of the triplets $(h,l,t)$, with $h,t \in E$, $l \in L$. Also, the embeddings $\mathbf{h,l,t}$ obtained values in $\mathbb{R}^{k}$ (i.e., $k$ is a hyperparameter) using the transformer-based number encoder and are denoted with the same letters in bold characters. In addition, the $L_1$- or $L_2$-norm can be used for the similarity measure $d(\cdot)$. To ensure the effectiveness of $\mathcal{L}_{\mathrm{KGE}}$, $\mathbf{l}$ is represented by a matrix filled with ones, and its parameters are not trained. Finally, the regression controller model's goal is to jointly optimize the following contrast and CN-KG embedding loss functions:

\begin{equation}
\mathcal{L} = \alpha \mathcal{L}_{\mathrm{NCE}} +(1-\alpha) \mathcal{L}_{\mathrm{KGE}},
\end{equation}

\noindent where $\alpha$ represents the joint optimization weight adjustment factor for the two loss functions. 

% Due to the small amount of task-specific data, the difficulty of the comparative learning task is low, and the training cycle of the regression controllers is short, which can be quickly adapted to different tasks.

% On this basis, in order to enable the text module's ability to represent the numerical labels of relationships at a standard level of ordinal cognition, we introduce a numerical commonsense knowledge graph to enhance the ordinal representation of numerical texts. Due to the small amount of task-specific data and the low difficulty of the comparison learning task, the training period of the regression controller is short which can quickly be adapted to different tasks.

% The regression controller is the guidance of the molecule generation model. Compared with classifier guidance and classifier-free guidance, regression controller can effectively perceive regression labels. This is one of the classifier-guided methods.

% The input to this module is the molecule and the regression label of that molecule in the corresponding task, as well as the eigenvector of another object. Use the following templates to generate the corresponding natural language text.

\subsubsection{The dual-branch controlled noise prediction model}

The limited diversity of molecular distributions in a specific task may lead to the poor performance of noise prediction networks trained on small molecular datasets. To ensure that the network adapts to the molecular and conditional distributions in a new task while remembering the prior knowledge, we proposed the dual-branch controlled noise prediction model, named DBControl. The DBControl model consists of two structurally consistent GNNs (i.e., the two networks denoted as $B_1$ and $B_2$, detailed structural diagrams can be found in the Appendix \ref{fig:DControl-NP}). $B_1$ participates in unconditional molecule training while $B_2$ does not. During the conditional mixed training phase, $B_2$ obtains its weights from $B_1$ for encoding conditional features. Notably, $B_1$ and $B_2$ weights must be optimized simultaneously, instead of simply freezing $B_1$ \citep{controlnet}. 

% (including target $\mathrm{IC}_{50}$ score $\mathbf{X_{number}}$ and target cell line $\mathbf{X_{cell}}$)

We represented conditional information $\mathbf{c}$ via $\Phi_{\mathrm{cap}}$ to obtain $\bm{C})$ (see Section \ref{sec:regcontroller}). For time $t$, two DBControl models $B_{\phi}$ = (${B}_{\phi, 1}$,${B}_{\phi, 2}$) and $B_{\theta}$ = (${B}_{\theta, 1}$,${B}_{\theta, 2}$) were proposed to estimate scores $\nabla_{\bm{A}}\mathrm{log}p_{t}(\bm{X}_t, \bm{A}_t, \bm{C}_t)$ and $\nabla_{\bm{X}}\mathrm{log}p_{t}(\bm{X}_t, \bm{A}_t, \bm{C}_t)$, respectively, expressed as follows:

\begin{equation}
\left\{
\begin{aligned}
\bm{X}^{l+1}&,\bm{A}^{l+1}\!=\! B_{\phi, 1}^{l}(\bm{X}^{l},\bm{A}^{l})+ \mathcal{Z}  (B_{\phi, 2}^{l}(\bm{X}^{l}+\mathcal{Z}  (\bm{C}^{l})-\overline{\mathcal{Z}  }(\bm{C}^{l-1}),\bm{A}^{l}+\mathcal{Z}  (\bm{C}^{l})-\overline{\mathcal{Z}  }(\bm{C}^{l-1}))) \\
\bm{X}^{l+1}&\!=\! B_{\theta, 1}^{l}(\bm{X}^{l},\bm{A}^{l})+ \mathcal{Z}  (B_{\theta, 2}^{l}(\bm{X}^{l}+\mathcal{Z}  (\bm{C}^{l})-\overline{\mathcal{Z}  } (\bm{C}^{l-1}+\delta(t)),\bm{A}^{l}))
\end{aligned}
\right.
\end{equation}

% \begin{equation}
% \left\{
% \begin{aligned}
% \bm{X}^{l+1} &=& & B_1^{l}(\bm{X}^{l},\bm{A}^{l})+ \mathcal{Z}  (B_2^{l}(\bm{X}^{l}+\mathcal{Z}  (c^{l}+\delta(t)),\bm{A}^{l}+\mathcal{Z}  (c^{l}+\delta(t)))), &l=1 \\
% \bm{X}^{l+1} &=& & B_1^{l}(\bm{X}^{l},\bm{A}^{l})+ \mathcal{Z}  (B_2^{l}(\bm{X}^{l}+\mathcal{Z}  (c^{l})-\mathcal{Z}  (c^{l-1}),\bm{A}^{l}+\mathcal{Z}  (c^{l})-\mathcal{Z} (c^{l-1}))), &l>1
% \end{aligned}
% \right.
% \end{equation}

\noindent where $\mathcal{Z} (\cdot)$ and $\overline{\mathcal{Z} }(\cdot)$ are zero convolutional layers, and $\bm{C}^{l}=0$  when $l=1$ (where $l$ represents the number of GNN layers). 

% Two DBControl models can be denoted as $ B_{\phi},B_{\theta} = (B_{\phi,1},B_{\phi,2}),(B_{\theta,1},B_{\theta,2})$.

% We introduce more than one $\mathcal{Z} (\cdot)$ to ensure the network retains its memory of the unconditional molecular distributions during the fine-tuning process.

% \subsubsection{ } 

% [Diffusion models beat gans on image synthesis] 

% 训练的CLIP模型可以用来引导扩散到目标提示。 例如，Dhariwal和Nichol[11]使用对噪声图像进行预训练的分类器来引导生成到目标类。 在药物生成领域，生成特定的类别不足以满足研究，而且实际上类别引导分子生成并不是十分有效。为此，我们提出了Regression text guidance diffusion，生成与特定物体具有回归值的关系的分子。

% 具体来说，a作为特定对象的名称，b是任务的回归标签（参考在第3.3小节），利用ab的输入条件引导分子生成。我们没有单独训练回归器模型，类似classifier-free的工作，选择训练通过分数估计器εθ(zλ)参数化的无条件去噪扩散模型pθ(z)和通过εθ(zλ，c)参数化的条件模型pθ(z|c)。不同的是，我们使用单个神经网络对单个模型进行参数化。单个模型分别处理有条件的输入和无条件的输入，防止模型遗忘和过拟合到特定任务数据。由于有无条件的数据混合训练，无条件的数据会在文本和数值编码时均标记为空文本而不是0.这是因为0在我们的框架中也是有意义的一个数据。
% 采样时使用条件和无条件的线性组合计算分数估计:

% The CLIP \citep{CLIP} can be used to guide diffusion to the target class.

\textbf{Regressor-free guidance diffusion.} Classifier guidance diffusion methods are typically insufficient for the DRP task (the algorithm is referenced in Algorithm \ref{alg:training}). Hence, we proposed the regressor-free guidance molecule generation method to support DRP. To guide the generation process towards the desired conditioning information $\mathbf{c}$ for sampling, one can sample from a conditional distribution $q_0(\bm{G}|\mathbf{c})$.  The expectations are carried over to the samples $\bm{G}_0\sim p_{data}$ and $\bm{G}_t\sim p_{0t}(\bm{G}_t|\bm{G}_0,\mathbf{c})$. Therefore, the transition probability $p_{0t}(\bm{G}_{t}|\bm{G}_{0},\mathbf{c})$ can be represented as follows:

\begin{equation}
    p_{0t}(\bm{G}_{t}|\bm{G}_{0},\mathbf{c}) = p_{0t}(\bm{X}_{t}|\bm{X}_{0},\mathbf{c})p_{0t}(\bm{A}_{t}|\bm{A}_{0},\mathbf{c}).
\end{equation}

Then, to minimize the Euclidean distance, we introduced the objectives that generalize the score matching \citep{songyang} to estimate the scores as follows:
% is approximated using the DBControl model $B_{\phi,\theta}$ of the form:

\begin{equation}
    \label{eq:expectations}
    \begin{array}{lc}
    \underset{\theta }{\mathrm{min} } \mathbb{E}_t\left \{ \lambda _1(t)\mathbb{E}_{\bm{G}_0}\mathbb{E}_{\bm{G}_t|\bm{G}_0}\left \|  B_{\theta,t }(\bm{G}_t,\mathbf{c})- \nabla_{\bm{X}_t} \mathrm{log} \; p_{0t}\left ( \bm{X}_t|\bm{X}_0,\mathbf{c} \right ) \right \|^2_2   \right \}, \\
    \underset{\phi }{\mathrm{min} } \mathbb{E}_t\left \{ \lambda _2(t)\mathbb{E}_{\bm{G}_0}\mathbb{E}_{\bm{G}_t|\bm{G}_0}\left \|  B_{\phi,t }(\bm{G}_t, \mathbf{c})- \nabla_{\bm{A}_t} \mathrm{log} \; p_{0t}\left ( \bm{A}_t|\bm{A}_0,\mathbf{c} \right ) \right \|^2_2   \right \},
    \end{array}
\end{equation}

\noindent where $\lambda_1(t)$ and $\lambda_2(t)$ are positive weighting functions. The expectations in Eq. \ref{eq:expectations} can be efficiently computed using the Monte Carlo estimate with the samples $(t, \mathbf{c}, \bm{G}_0, \bm{G}_t)$ \citep{GDSS}.

\textbf{Regressor-free guidance sampling}. Let $\bm{G} \gets (\bm{X} \sim \mathcal{N}(\bm{0}, \bm{I}), \bm{A} \sim \mathcal{N}(\bm{0}, \bm{I}) )$, $\mathbf{z} = \left \{ \mathbf{z}_\lambda \mid \lambda \in \left [ \lambda_{min} ,\lambda_{max}  \right ]   \right \} $ for hyperparameters $\lambda_{min} < \lambda_{max} \in \mathbb{R} $, which $\mathbf{z}_\lambda = \alpha_\lambda \bm{G}+\sigma_\lambda\epsilon $. Then, trained DBControl models $B_{\phi}$ and $B_{\theta}$ can simulate the system of reverse-time SDEs Eq.\ref{eq:reverse-time-SDEs} to enable the reverse-time diffusion process as a generative model. The two DBControl models used for estimating true scores $\nabla _{\bm{X}}\mathrm{log} \; p_t(\bm{X}|\mathbf{c} )$ and $\nabla _{\bm{A}}\mathrm{log} \; p_t(\bm{A}|\mathbf{c} )$ respectively are abbreviated as $\bm{\epsilon_\theta}(\mathbf{z}_\lambda,\mathbf{c})$. In the conditional mixed training phase, when unconditional data are the input,  $\mathbf{c} $ are labeled as empty text $\emptyset$, denoted as $\bm{\epsilon_\theta}(\mathbf{z}_\lambda,\emptyset)$. Based on the Eq. \ref{eq2}, the score estimation is calculated using a conditional and unconditional linear combination when sampling:

% parameterized by $(\bm{\epsilon_\theta}^{\bm{X}} (\mathbf{z}_\lambda|\mathbf{c}),\bm{\epsilon_\theta}^{\bm{A}} (\mathbf{z}_\lambda|\mathbf{c}))$. 

% We use a single neural network to parameterize both denoising diffusion models, handling conditional and unconditional inputs separately to prevent the model from forgetting or overfitting task-specific data. 

\begin{equation}
\widetilde{\bm{\epsilon}}_{\bm{\theta}}(\mathbf{z}_\lambda,\mathbf{c})=(1+w)\bm{\epsilon_\theta}(\mathbf{z}_\lambda, \mathbf{c})-w\bm{\epsilon_\theta}(\mathbf{z}_\lambda,\emptyset ),
\end{equation}

\noindent where $w$ is a conditional control strength parameter ($w>0$), and $w=0$ indicates unconditional generation. $\widetilde{\bm{\epsilon}}_{\bm{\theta}}(\mathbf{z}_\lambda,\mathbf{c})$ is then used in place of $\bm{\epsilon_\theta}(\mathbf{z}_\lambda, \mathbf{c})$ when sampling, resulting in approximate samples from the distribution $\widetilde{p}_t (\mathbf{z}_\lambda,\mathbf{c} ) \propto {p}_t (\mathbf{z}_\lambda,\mathbf{c} ){p}_t (\mathbf{c},\mathbf{z}_\lambda )^w$. The sampling algorithm is referenced in Algorithm \ref{alg:em-solver}.

% Notably,  $\mathbf{c}$  is not randomly set as empty; instead, we combined conditional and unconditional datasets and jointly trained the DBControl model, making the probability constant.

% (randomly discarding information from the conditional dataset, which has very few samples, would be unreasonable).

% training
\begin{algorithm}[!ht]
\caption{Joint training the DBControl model with regressor-free guidance}
\label{alg:training}
\textbf{Require}:  molecular data for the specific task  $\left \{ \bm{G}_0=(\bm{X}_0,\bm{A}_0), \mathbf{c} \right \} $, the DBControl models $\bm{\epsilon_\theta}$, condition embedding model $\bm{f}_\theta$, schedule function $\delta(\cdot)$, $\alpha(\cdot)$ and $\sigma(\cdot)$

\begin{algorithmic}[1]
\Repeat
    \State Sample $t \sim \mathcal{U}(0,1], \bm{\epsilon}_{\bm{X}} \sim \mathcal{N}(\bm{0}, \bm{I}), \bm{\epsilon}_{\bm{A}} \sim \mathcal{N}(\bm{0}, \bm{I})$ 
    \State $\bm{T}_0 \gets \varnothing, \bm{N}_0 \gets \varnothing $ with definite probability
    \State $\bm{C}_t = \delta(t)\bm{f}_\theta(\mathbf{c})$ \Comment{Embedding the regression values as conditions}
    \State $\bm{G}_t = (\bm{X}_t,\bm{A}_t)  \gets (\alpha(t) \bm{X}_0 + \sigma(t) \bm{\epsilon}_{\bm{X}}, \alpha(t) \bm{A}_0 + \sigma(t) \bm{\epsilon}_{\bm{A}})$ 
    \State $\bm{\epsilon}_{\bm{\theta}}^{\bm{X}}, \bm{\epsilon}_{\bm{\theta}}^ {\bm{A}} \gets \bm{\epsilon}_{\bm{\theta}}(\bm{G}_t, \bm{A}_t,\bm{C}_t, t)$
    \Comment{Regressor-free guidance generation}
    \State Minimize $||\bm{\epsilon}_{\bm{\theta}}^{\bm{X}} - \bm{\epsilon}_{\bm{X}}||^2_2 + 
    ||\bm{\epsilon}_{\bm{\theta}}^ {\bm{A}} - \bm{\epsilon}_{\bm{A}}||^2_2$
\Until{converged}
\end{algorithmic}
\end{algorithm}

% sampling with EM solvers
\begin{algorithm}[!ht]
\caption{Conditional sampling with regressor-free guidance
}
\label{alg:em-solver}
\textbf{Require}: number of time steps $N$, the DBControl models $\bm{\epsilon_\theta}$, drift coefficient function $f(\cdot)$, diffusion coefficient function $g(\cdot)$, schedule function $\sigma(\cdot)$, post-processing function $post(\cdot)$,  conditioning information for conditional sampling $\mathbf{c}$, condition embedding model $\bm{f}_\theta$

\begin{algorithmic}[1]
    \State Sample initial graph $\bm{G} \gets (\bm{X} \sim \mathcal{N}(\bm{0}, \bm{I}), \bm{A} \sim \mathcal{N}(\bm{0}, \bm{I}) )$
    \State $\Delta t = \frac{T}{N}$
    \For{$i \gets N$ to $1$}
    \State $\bm{\epsilon}_{\bm{X}} \sim \mathcal{N}(\bm{0}, \bm{I}), \bm{\epsilon}_{\bm{A}} \sim \mathcal{N}(\bm{0}, \bm{I})$
    \State $t \gets i\Delta t$
    \State $\bm{C} = \delta(t)\bm{f}_\theta(\mathbf{c})$ \Comment{Embedding the regression values as conditions}
    \State $\bm{\epsilon}_{\bm{\theta}}^{\bm{X}}, \bm{\epsilon}_{\bm{\theta}}^ {\bm{A}} \gets (1+w)\bm{\epsilon}_{\bm{\theta}}(\bm{G}, \bm{A}, \bm{C}, t)-w\bm{\epsilon}_{\bm{\theta}}(\bm{G}, \bm{A}, \varnothing , t)$
    \Comment{Regressor-free guidance sampling}
    \State $\bm{X} \gets \bm{X} - (f(t)\bm{X}+\frac{g(t)^2}{\sigma(t)}\bm{\epsilon}_{\bm{\theta}}^{\bm{X}})\Delta t + g(t)\sqrt{\Delta t} \bm{\epsilon}_{\bm{X}}$
    \State $\bm{A} \gets \bm{A} - (f(t)\bm{A}+\frac{g(t)^2}{\sigma(t)}\bm{\epsilon}_{\bm{\theta}}^{\bm{A}})\Delta t + g(t)\sqrt{\Delta t} \bm{\epsilon}_{\bm{A}}$
    \EndFor
    \State \Return $post(\bm{X}, \bm{A})$
\end{algorithmic}
\end{algorithm}

\section{Theoretical discussion}
\label{Alg}
In this section, we first address why the regressor-free guidance sampling method is more effective than the classifier-based guidance sampling for molecules. Second, we aim to demonstrate the regression controller model's effectiveness and how it accurately guides the diffusion model to perceive different $\mathrm{IC}_{50}$ scores when it is used as a condition. The $\mathrm{IC}_{50}$ scores are represented as continuous values within the range of 0 to 1 after normalization.

\noindent \textbf{Notations}. Let the classifier's class threshold be denoted as $T \in \left ( 0,1 \right ) $, where the conditional input is $\left \{ 0, 1 \right \}$. The conditional input for the regressor is $(0,1)$, and the target $\mathrm{IC}_{50}$ score is denoted as $\bm{C}_{\mathrm{aim} } \in \mathbb{R}$. The classifier and regressor's errors during supervised training are denoted as $\varepsilon_1$ and $\varepsilon_2$, respectively. Their sampling spaces are $\bm{S}_{cls} = \left [0 \pm  \varepsilon_1  ,T \pm  \varepsilon_1 \right ]$ and $\bm{s}_{reg} = \left [ \bm{C}_{\mathrm{aim} }-\frac{10^{-\xi}}{2} \pm\varepsilon_2,\bm{C}_{\mathrm{aim} } +\frac{10^{-\xi}}{2} \pm\varepsilon_2 \right ] $, where $0< \bm{C}_{\mathrm{aim} } < T < 1$. $0$ and $T$ are the upper and lower bounds of the category corresponding to the contained $\bm{C}_{\mathrm{aim} }$. In addition, the sampling space size is denoted as $\left \| \bm{S} \right \|$. Then, the two sampling space sizes can be represented as $\left \| \bm{S}_{cls} \right \| $, $ \left \| \bm{S}_{reg} \right \|  $.

% = T_2-T_1 +2 \varepsilon_1 
% =2\varepsilon_2

\begin{proposition}[Main proposition]
    \label{thm:main}
    For any $\bm{C}_{\mathrm{aim} } \in\left ( 0,1 \right ) $,  then $\left \| S_{cls} \right \| \ge \left \| S_{reg} \right \|$  exists.
\end{proposition}

Proposition \ref{thm:main} states that the regression point sampling space is smaller than that of the class range (refer to Appendix Proof.\ref{proof_lemma1} for detailed proof). The two ranges are equal if $T = 10^{-\xi} $. Practically, $\xi$
is related to the model's performance. If $T \le 10^{-\xi} $, the  $\left ( 0, T \right ) $ collapses to a single point, becoming a  regressor-free guidance exception.
% \begin{equation}
%     \left \| S_{cls} \right \|  = T_2-T_1 +2 \varepsilon_1 , \;\;\; \left \| S_{reg} \right \|  =2\varepsilon_2
% \end{equation}

Second, we used a regression controller model to convert $\mathrm{IC}_{50}$ into text labels. However, we must demonstrate whether this method can effectively represent $\bm{C}_{\mathrm{aim} }$. $\bm{C}_{\mathrm{aim} }$ being effectively represented mathematically means that its representation is unique. Let the number encoder of the regression controller model (see Section \ref{sec:regcontroller}) be denoted as  $\Theta\left ( \cdot \right ) $.

\begin{proposition}[Uniqueness of $\bm{C}_{\mathrm{aim} }$ Representation]
    \label{lemma:unique}
   For any $\lambda \in \left [ 0,1 \right ] $,  and  $\lambda \ne \bm{C}_{\mathrm{aim} }$, then we say that  
   $\Theta(\bm{C}_{\mathrm{aim} })\ne \Theta(\lambda )$.
\end{proposition}

Proposition \ref{lemma:unique} indicates the uniqueness of $\bm{C}_{\mathrm{aim} }$ representation via the number encoder $\Theta\left ( \cdot \right ) $ (refer to Appendix Proof.\ref{proof_lemma2} for detailed proof). $E$ is the entity set of CN-KG and $L$ is the relationship called "is less than". We mapped $\mathrm{IC}_{50}$ to $E$ following the definition of a well ordered set \citep{wo_1}. Specifically, given that the $\mathrm{IC}_{50}$ scores in the dataset have a finite decimal precision of $\xi \in \mathbb{N}$, we constructed a well-ordered set $W$ of size $n$, where $n =10^{\xi}$, $n\in \mathbb{N}^{*}$:

\begin{equation}
    W[0]=0,\;W[1]=1,\;...,\;W[i]=i,\;W[n]=n,
\end{equation}

\noindent where $i\in [0,n]$. Therefore, each element $W[i]$ is unique, implying that in $W$, every $\mathrm{IC}_{50}$ is also unique and corresponds to its representation with precision $\xi$. Since $\Theta(W[i])$'s representation is constrained by $\mathcal{L}_{\mathrm{KGE}}$, we can construct a well-ordered set $V$:

\begin{proposition}[Equal interval representation of $\Theta$]
\label{thm:Equal_Interval}
   For any $\xi$,  a perturbation $\varepsilon_3 $ exist to make $ \Theta(V[i])-\Theta(V[i+1])  = \mathbf{l} + \varepsilon_3$.
\end{proposition}

\noindent where $V[i] = \Theta(W[i])$ and $\epsilon_3 \in \mathbb{R}$ is a model perturbation. $\mathbf{l}$ is the only relationship in CN-KG, represented as a matrix filled entirely with ones (i.e., $\left \| \mathbf{l}  \right \| =1$), which is not involved in the training. Proposition \ref{thm:Equal_Interval} states that $V$ is a well-ordered set under ideal conditions (i.e., $\varepsilon_3$ near to 0, refer to Appendix Proof.\ref{proof_lemma3} for detailed proof). As a result, Proposition \ref{lemma:unique} can be proved due to the uniqueness of the elements in the well-ordered set (See Appendix \ref{proof_lemma2} for details). Therefore, the regression controller model can effectively represent any $\mathrm{IC}_{50}$.

% Secondly, the $\mathrm{IC}_{50}$ score range is a continuous interval, and $\bm{C}_{\mathrm{aim} }$ should be effectively represented. The Regression controller proposed in this paper serves as a label encoder, utilizing CN-KG to constrain the representation of label nodes within the interval. Specifically, $\ne \varnothing $

\section{Experiments}

\subsection{Experimental setup}

We selected the DRP task's GDSCv2 as the conditional molecular dataset and QM9 as the unconditional dataset (refer to Table \ref{tab:mol_dataset_uncondition} and Table \ref{tab:mol_dataset_condition}) for resource consideration. The evaluation metrics include Fréchet chemNet distance (FCD) \citep{FCD} and Neighborhood subgraph pairwise distance kernel maximum mean discrepancy (NSPDK MMD) \citep{MMD}. Further details regarding the experimental environment and hyperparameter tuning are detailed in the Appendix \ref{D}. Error experiments for the method are referenced as Appendix \ref{apx:ErrorAssessment}, while visualization-related information is referenced as Appendix \ref{Visualization}.

\subsection{Overall experiment}
\label{exp:overall}
% 我们展示最主要的实验，通过我们的模型生成的分子是否能够感知回归标签并生成特定条件下的分子。采用分子生成任务中，两个最常用的衡量指标，fcd和mmd。我们对比了近几年分子生成主流方法。表1和表二分别展示了。两个指标下，我们的模型和主流分子生成模型的性能。

We evaluated whether the molecules generated by our model can accurately predict regression labels and generate molecules under specific conditions. Then, we compared our method with representative molecule generation methods (refer to 
Appendix \ref{A_Baselines} for further details). Tables \ref{tab:x1} and \ref{tab:x2} display the performance comparison between our model and the mainstream molecule generation models, with the metrics being FCD and MMD. 

\begin{figure}[h!]
    \centering
    \begin{subfigure}[b]{0.235\textwidth}
        \centering
        \includegraphics[width=\textwidth]{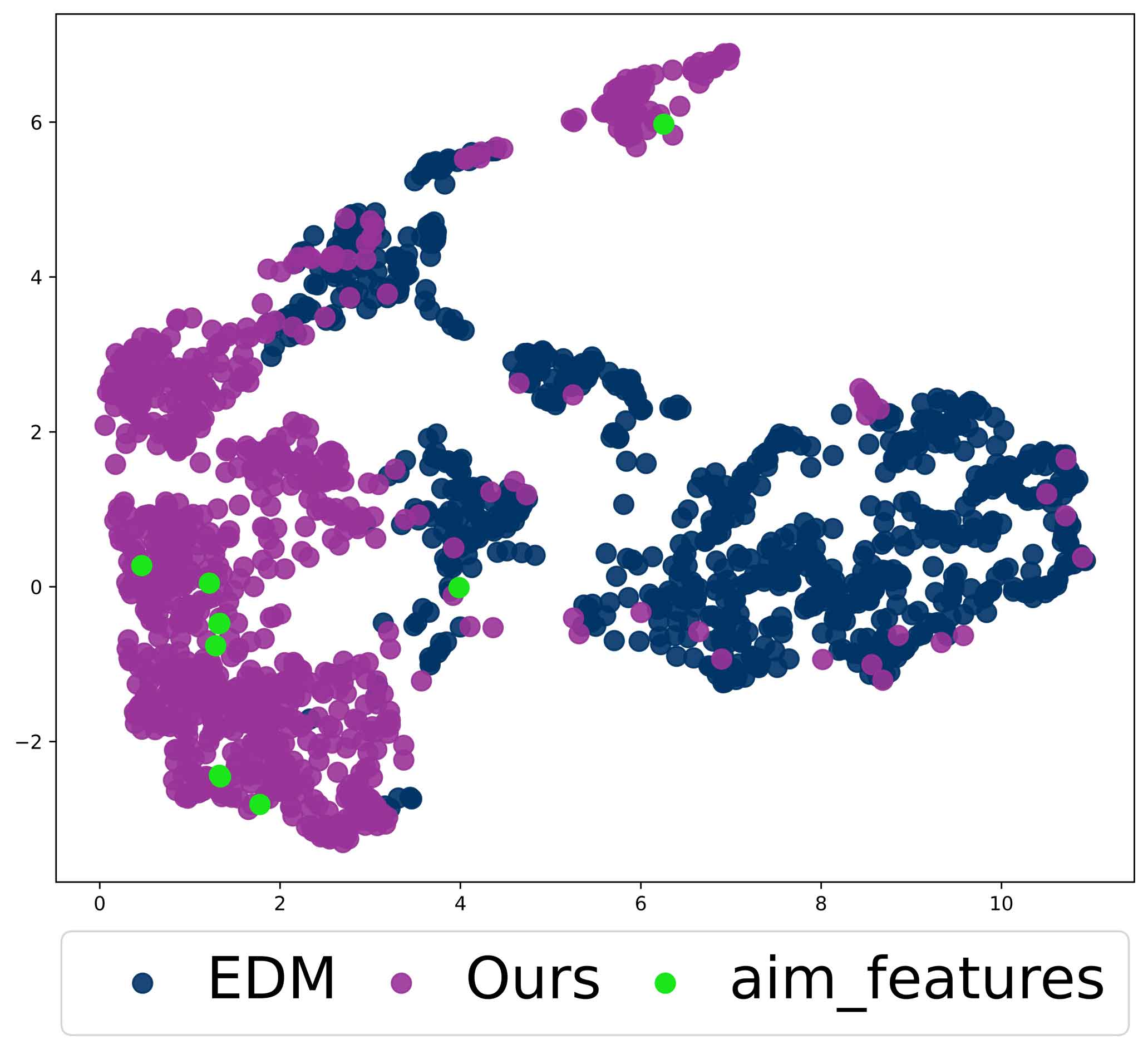}
        \caption{EDM}
        \label{fig:image1}
    \end{subfigure}
    \begin{subfigure}[b]{0.235\textwidth}
        \centering
        \includegraphics[width=\textwidth]{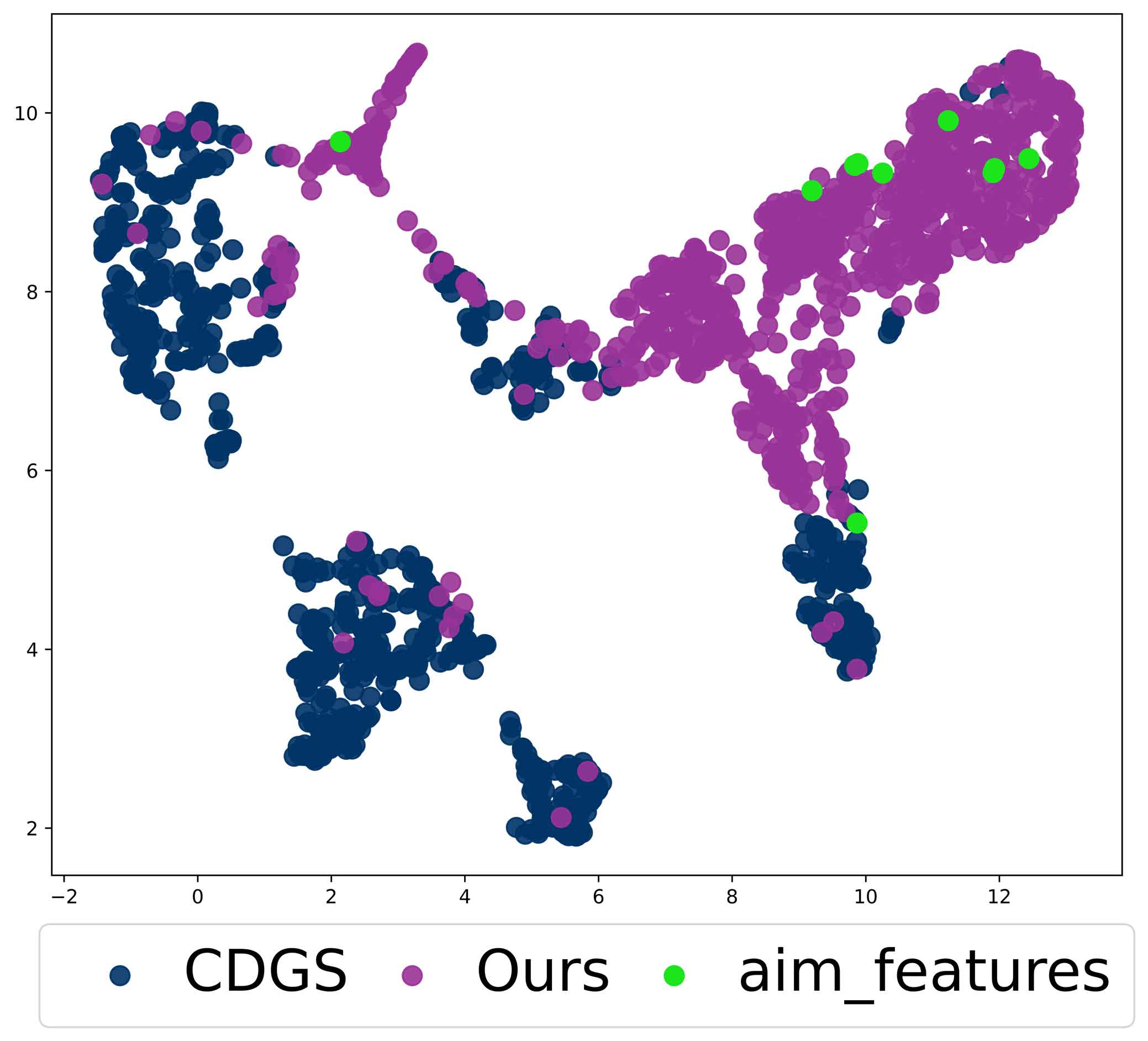}
        \caption{CDGS}
        \label{fig:image2}
    \end{subfigure}
    \begin{subfigure}[b]{0.235\textwidth}
        \centering
        \includegraphics[width=\textwidth]{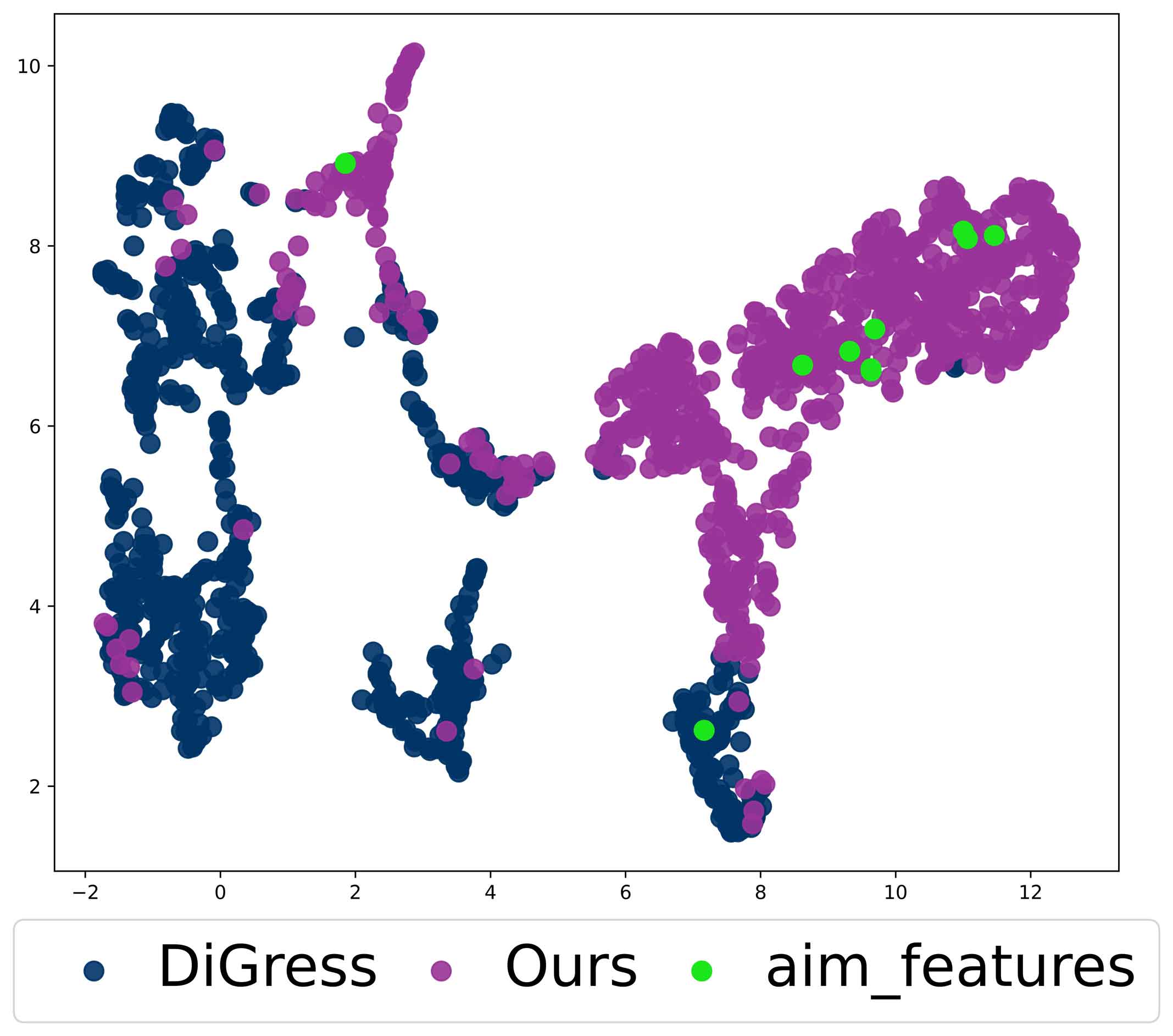}
        \caption{DiGress}
        \label{fig:image3}
    \end{subfigure}
    \begin{subfigure}[b]{0.235\textwidth}
        \centering
        \includegraphics[width=\textwidth]{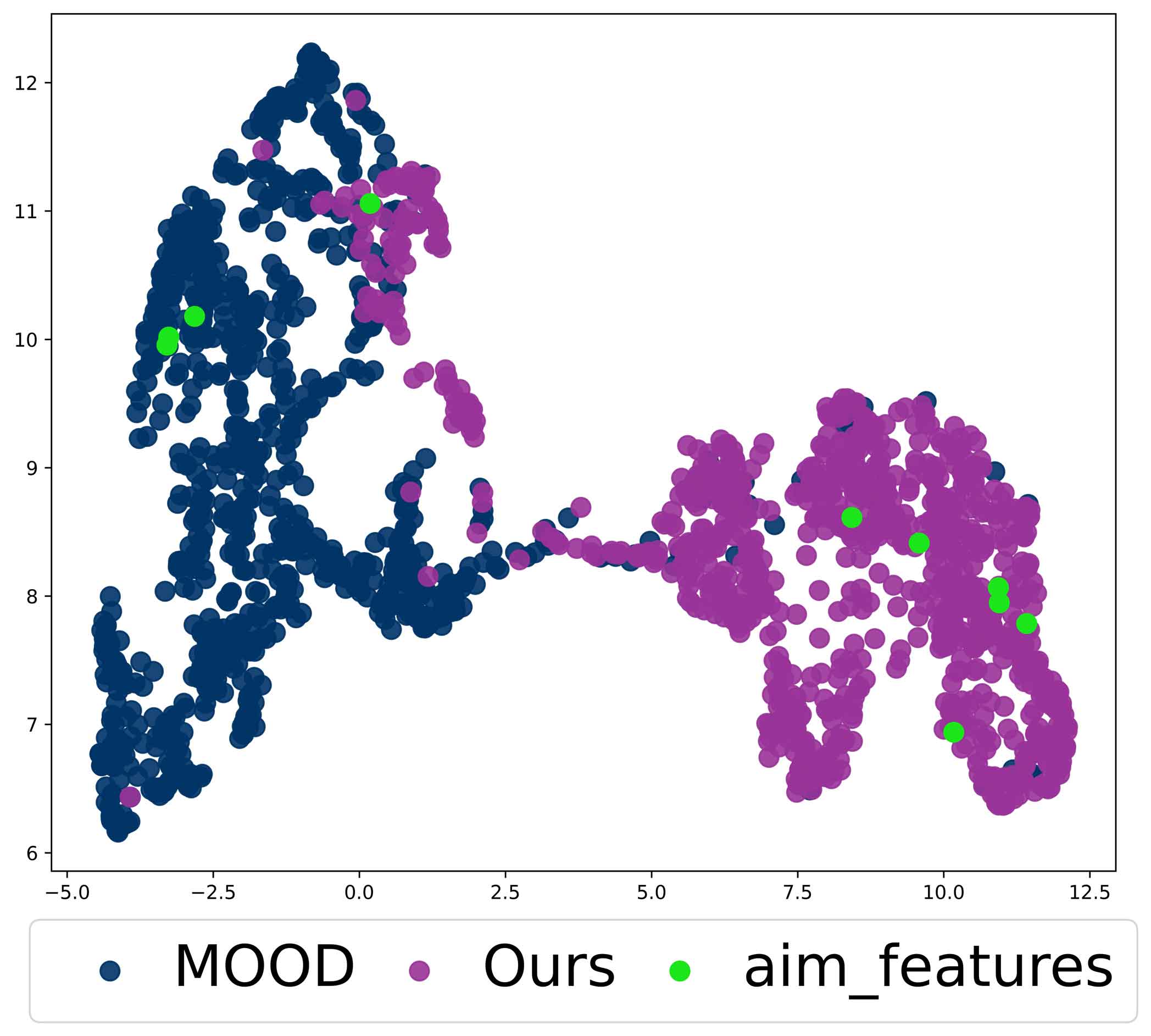}
        \caption{MOOD}
        \label{fig:image4}
    \end{subfigure}
    \caption{UMAP visualization of molecule generation results with our method compared to four mainstream methods using the target pair (NCI-H187, $\mathrm{IC}_{50}$=0.35).}
    \label{fig:four_images}
\end{figure}

The TopK refers to the top $K$ molecules that meet the criteria of a specific cell line and $\mathrm{IC}_{50}$ score. When $K$ is set to 1, the target molecule is the one that reacts closest to the target $\mathrm{IC}_{50}$ for a specific cell line. Therefore, a smaller $K$ indicates greater task difficulty. We trained and sampled the recent methods. The results indicate that our model achieved the best performance in all metrics and cell line tasks. Specifically, the FCD and MMD were 2.68\% and 2.23\% higher than the best model, respectively. Furthermore, numerical values cannot intuitively demonstrate our method's superiority. Therefore, we selected four mainstream methods and visualized a set of generation data for the target pair (NCI-H187, $\mathrm{IC}_{50}$=0.35). As shown in Fig. \ref{fig:four_images}, the molecules generated by our method are mainly centered around condition sampling, while those from other methods deviate significantly from the target value.

% \usepackage{graphicx}
\begin{table}[!t]
\centering
\caption{\textbf{Results of FCD}. The metric is calculated with 1000 samples generated from each model with the TopK (K=3 / 5 / 10 / 15 / 20).  A lower number indicates a better generation quality, and the best performance is highlighted in bold. }
\renewcommand{\arraystretch}{1.45}

\label{tab:x1}
\setlength{\tabcolsep}{2mm}{
\resizebox{\textwidth}{!}{%
\begin{tabular}{lcccc}
\hline
Method & ES3, $\mathrm{IC}_{50}$=0.4 & ES5, $\mathrm{IC}_{50}$=0.4 & NCI-H187, $\mathrm{IC}_{50}$=0.35 & Hs-578-T, $\mathrm{IC}_{50}$=0.4 \\ \hline
GDSS \citep{GDSS} & 83.2\>/\>77.4\>/\>60.2\>/\>56.2\>/\>52.1 & 72.6\>/\>65.6\>/\>58.9\>/\>55.2\>/\>58.9 & 95.1\>/\>92.5\>/\>83.8\>/\>80.2\>/\>74.0 & 104.5\>/\>74.8\>/\>73.4\>/\>70.0\>/\>65.3 \\
EDM \citep{E3Diffusion} & 82.9\>/\>76.3\>/\>59.8\>/\>56.7\>/\>52.9 & 72.4\>/\>65.5\>/\>59.3\>/\>56.3\>/\>59.9 & 94.4\>/\>91.7\>/\>83.6\>/\>80.4\>/\>74.6 & 104.5\>/\>75.5\>/\>73.8\>/\>71.0\>/\>66.3 \\
CDGS \citep{CDGS} & 85.0\>/\>77.8\>/\>61.1\>/\>57.4\>/\>53.1 & 74.1\>/\>66.7\>/\>59.6\>/\>56.3\>/\>60.2 & 95.4\>/\>93.1\>/\>84.9\>/\>81.1\>/\>74.8 & 106.6\>/\>76.7\>/\>74.4\>/\>71.0\>/\>66.5 \\
GeoLDM \citep{GeoLDM} & 84.2\>/\>78.2\>/\>61.3\>/\>57.7\>/\>53.7 & 73.9\>/\>66.9\>/\>60.2\>/\>56.7\>/\>60.2 & 95.4\>/\>92.9\>/\>84.7\>/\>81.5\>/\>75.5 & 105.5\>/\>76.4\>/\>74.7\>/\>71.6\>/\>67.1 \\
DiGress \citep{DiGress}& 85.2\>/\>77.7\>/\>60.9\>/\>57.4\>/\>53.0 & 74.3\>/\>67.4\>/\>60.1\>/\>56.9\>/\>60.4 & 96.9\>/\>94.0\>/\>85.2\>/\>81.1\>/\>74.6 & 106.7\>/\>77.2\>/\>74.5\>/\>71.3\>/\>66.3 \\
MOOD \citep{MOOD}& 80.2\>/\>\textbf{71.0}\>/\>57.7\>/\>53.8\>/\>48.7 & 70.0\>/\>64.7\>/\>55.5\>/\>52.9\>/\>57.2 & 92.3\>/\>88.0\>/\>77.4\>/\>75.1\>/\>69.9 & 100.8\>/\>71.4\>/\>\textbf{67.5}\>/\>65.5\>/\>61.6 \\
GruM-2D \citep{GruM2D}& 85.3\>/\>77.4\>/\>60.7\>/\>57.1\>/\>52.7 & 73.5\>/\>66.5\>/\>59.6\>/\>56.3\>/\>59.7 & 97.2\>/\>94.2\>/\>84.9\>/\>80.4\>/\>74.4 & 105.5\>/\>76.3\>/\>73.9\>/\>71.0\>/\>66.3 \\ \hline
Ours(w=0) & 
78.6\>/\>73.0\>/\>62.4\>/\>60.5\>/\>56.4 & 69.2\>/\>\textbf{61.5}\>/\>57.5\>/\>56.3\>/\>59.8 & 92.5\>/\>88.1\>/\>81.9\>/\>79.3\>/\>74.4 & 99.9\>/\>72.3\>/\>72.6\>/\>70.2\>/\>66.7 \\
Ours(w=1) & 
\textbf{77.0}\>/\>72.3\>/\>\textbf{56.0}\>/\>\textbf{52.1}\>/\>\textbf{47.8} & \textbf{68.8}\>/\>63.3\>/\>\textbf{53.6}\>/\>\textbf{50.7}\>/\>\textbf{54.9} & \textbf{90.2}\>/\>\textbf{86.2}\>/\>\textbf{76.1}\>/\>\textbf{74.1}\>/\>\textbf{68.5} & \textbf{98.3}\>/\>\textbf{68.8}\>/\>67.7\>/\>\textbf{64.3}\>/\>\textbf{60.1} \\ \hline
\end{tabular}%
}}
\end{table}

% \usepackage{graphicx}
\begin{table}[!t]
\centering
\caption{\textbf{Results of MMD}. The remaining setting is the same as in Table \ref{tab:x1}.}

% The metric is calculated with 1000 samples generated from each model. The results are the TopK (K=3 / 5 / 10 / 15 / 20).  A lower number indicates a better generation quality, and the best performance is highlighted in bold.

\renewcommand{\arraystretch}{1.45}
\label{tab:x2}
\resizebox{\textwidth}{!}{%
\begin{tabular}{lcccc}
\hline
Method & ES3, $\mathrm{IC}_{50}$=0.4 & ES5, $\mathrm{IC}_{50}$=0.4 & NCI-H187, $\mathrm{IC}_{50}$=0.35 & Hs-578-T, $\mathrm{IC}_{50}$=0.4 \\ \hline
GDSS \citep{GDSS} & .351\>/\>.267\>/\>.150\>/\>.129\>/\>.119 & .401\>/\>.277\>/\>.177\>/\>.146\>/\>.133 & .433\>/\>.283\>/\>.187\>/\>.163\>/\>.145 & .338\>/\>.232\>/\>.177\>/\>.145\>/\>.134 \\
EDM \citep{E3Diffusion} & .338\>/\>.251\>/\>\textbf{.139}\>/\>.116\>/\>.106 & .382\>/\>.262\>/\>.165\>/\>.134\>/\>.120 & .411\>/\>.265\>/\>.172\>/\>.148\>/\>.130 & .324\>/\>.222\>/\>.163\>/\>.132\>/\>.121 \\
CDGS \citep{CDGS} & .340\>/\>.255\>/\>.142\>/\>.118\>/\>.110 & .391\>/\>.269\>/\>.169\>/\>.137\>/\>.122 & \textbf{.410}\>/\>.264\>/\>\textbf{.170}\>/\>.149\>/\>.131 & .326\>/\>.225\>/\>.165\>/\>.134\>/\>.123 \\
GeoLDM \citep{GeoLDM} & .340\>/\>.255\>/\>.146\>/\>.121\>/\>.113 & .385\>/\>.265\>/\>.167\>/\>.137\>/\>.123 & .418\>/\>.269\>/\>.176\>/\>.152\>/\>.134 & .322\>/\>.223\>/\>.168\>/\>.138\>/\>.125 \\
DiGress \citep{DiGress} & .341\>/\>.255\>/\>.147\>/\>.122\>/\>.113 & .386\>/\>.265\>/\>.166\>/\>.137\>/\>.123 & .422\>/\>.272\>/\>.178\>/\>.154\>/\>.134 & .324\>/\>.223\>/\>.168\>/\>.137\>/\>.124 \\
MOOD \citep{MOOD} & .347\>/\>.242\>/\>.195\>/\>.152\>/\>.144 & \textbf{.304}\>/\>.229\>/\>.150\>/\>.139\>/\>.131 & .463\>/\>.311\>/\>.216\>/\>.174\>/\>.151 & .305\>/\>.226\>/\>.180\>/\>.160\>/\>.138 \\
GruM-2D \citep{GruM2D} & .337\>/\>.250\>/\>.138\>/\>.116\>/\>.106 & .386\>/\>.263\>/\>.164\>/\>.133\>/\>.120 & .410\>/\>\textbf{.263}\>/\>.172\>/\>.148\>/\>.130 & .325\>/\>.220\>/\>.163\>/\>.131\>/\>.121 \\ \hline
Ours(w=0) & .387\>/\>.300\>/\>.207\>/\>.184\>/\>.164 & .384\>/\>.258\>/\>.177\>/\>.159\>/\>.142 & .458\>/\>.297\>/\>.212\>/\>.180\>/\>.159 & .334\>/\>.244\>/\>.195\>/\>.164\>/\>.146 \\
Ours(w=1) & 
\textbf{.313}\>/\>\textbf{.221}\>/\>.142\>/\>\textbf{.113}\>/\>\textbf{.101} & .327\>/\>\textbf{.226}\>/\>\textbf{.135}\>/\>\textbf{.116}\>/\>\textbf{.107} & .428\>/\>.273\>/\>.178\>/\>\textbf{.146}\>/\>\textbf{.123} & \textbf{.299}\>/\>\textbf{.200}\>/\>\textbf{.158}\>/\>\textbf{.129}\>/\>\textbf{.109} \\ \hline
\end{tabular}%
}
\end{table}

\subsection{Varying the regressor-free guidance strength experiment}

% VARYING THE CLASSIFIER-FREE GUIDANCE STRENGTH

% The model is a model based on regression perception and relationships, generation conditions, and molecular diffusion. Therefore, it is necessary to sample the different regression values to explore whether the model has the ability of regression label perception.

% In the same drug molecule directed generation task, using different conditions to affect the hyperparameters will affect the sampling results. To this end, we conducted a hyperparameter experiment to explore the effect of this parameter on the sampling results, with long intervals ranging from zero to 15.

% 分子定向生成任务中，使用不同强度的条件参数会影响采样结果。为此，我们进行了一个回归器引导强度实验，将我们提出的无回归器指导应用于ES3细胞系的IC50为0.4的条件生成。在图2中，我们展示了我们的模型在不同指导强度w下的样本质量效果。K是TopK。w不等间隔地从0到10选取。随着条件引导的强度变大，分子生成效果逐步向K个目标分子的特征值附近转移，但是在w=1附近达到第一个极值。随后在1<w<3,生成效果会变差，但在w=5附近又会下降，达到第二个极值。此外，K越大时，目标分子数量越多，目标IC50的范围越广，FCD和MMD指标越小。因此，更小的K是一种更严苛的条件生成任务。总结来说，我们通过实验证实了本文的主要观点：即无回归器指导能够引导扩散模型生成定向条件的分子，并且在强度w为1和5左右达到两个极小值点。

\begin{figure}[!h]
    \centering
    \includegraphics[width=0.95\linewidth]{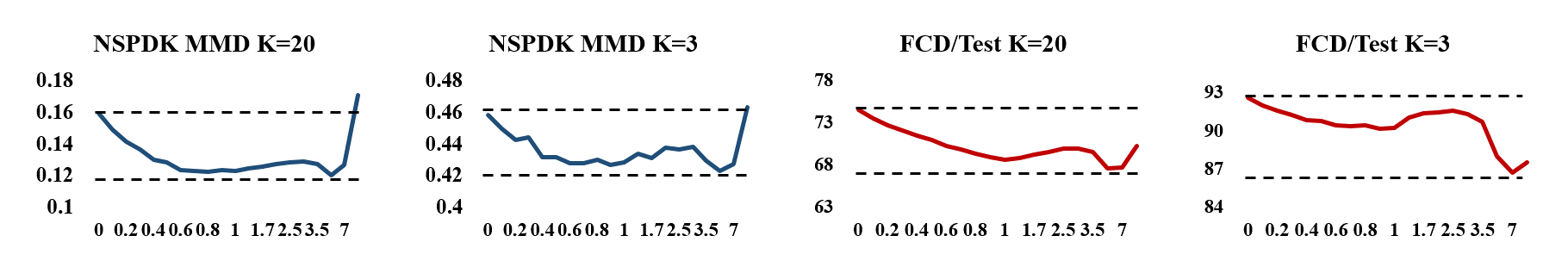}
    \caption{Visualization of regressor-free guidance strength trends. The x-axis represents the conditions' intensity,  where $w = 0.0$ refers to non-guided models, while the y-axis represents the corresponding metric values.}
    \label{fig:ablation_w_n}
\end{figure}

Different conditional parameters can affect the sampling results during conditional molecule generation tasks. To investigate this, we conducted a regressor-free guidance strength experiment, applying our proposed regression-free guidance to conditionally generate molecules with an $\mathrm{IC}_{50}$ of 0.4 for the cell line ES3 for resource consideration. In Fig. \ref{fig:ablation_w_n}, we demonstrate our model's sample quality effects under different guidance strengths $w$. In addition, $w$ was selected non-uniformly from 0 to 10. The visualization comparison images for molecule generation with conditional $w=1.0$ and unconditional guidance $w=0.0$ can be found in Appendix Fig. \ref{fig:umap}.

As the conditional guidance strength increases, the molecule generation process gradually transitions toward the $K$ target molecules feature values, but reaches the first extremal point at approximately $w = 1$. Subsequently, the generation performance deteriorates between $ 1 < w < 3 $, and improves at approximately $ w = 5 $, reaching the second extremal point. Additionally, as $ K $ increases, the number of target molecules and the range of target $\mathrm{IC}_{50}$ scores expand, while the FCD and MMD metrics decrease. Thus, a smaller $ K $ represents a more stringent conditional generation task. In summary, we experimentally verified this paper's main hypothesis: that regression-free guidance can steer generation models to generate directionally conditioned molecules, with two minimal guidance strength points $ w $ of 1 and 5.

\subsection{Ablation studies}

\begin{table}[!h]
\centering
\caption{Our method’s ablation experiments involve the target pair (NCI-H187, $\mathrm{IC}_{50}$=0.35) with TopK values of 3, 5, 10, 15, and 20. The optimal results are indicated in \textbf{bold}, while the second-best results are \underline{underscored}. The dash (-) signifies cases where both scenarios (not loading pre-trained models and unfreezing model weights) cannot occur simultaneously.}
\label{tab:ablation}
\renewcommand{\arraystretch}{1.5}
\resizebox{\textwidth}{!}{%
\begin{tabular}{ccccccc}
\hline  
\begin{tabular}[c]{@{}c@{}}Mixed data \\ training\end{tabular} & \begin{tabular}[c]{@{}c@{}}Uncondition\\pre-trained\end{tabular} & \begin{tabular}[c]{@{}c@{}}$B_1$\\freezing\end{tabular} & FCD & MMD & \begin{tabular}[c]{@{}c@{}}Validity\\ w/o correction\end{tabular} & Valid \\ \hline
\ding{55} & \ding{55}  & \ding{55}  & 82.2\>/\>77.7\>/\>62.0\>/\>57.4\>/\>54.1 & 0.344\>/\>0.258\>/\>0.168\>/\>0.141\>/\>0.133 & 0.637 & 1 \\
\ding{55}  & \ding{55}  & \ding{51}  & - & - & - & - \\
\ding{55}  & \ding{51}  & \ding{51}  & 83.3\>/\>76.4\>/\>64.5\>/\>63.0\>/\>60.2 & 0.490\>/\>0.412\>/\>0.288\>/\>0.270\>/\>0.254 & 0.000 & 1 \\
\ding{55}  & \ding{51}  & \ding{55}  & \textbf{75.9}\>/\>\textbf{71.3}\>/\>\textbf{54.9}\>/\>\textbf{51.0}\>/\>\textbf{46.8} & \underline{0.319}\>/\>\underline{0.232}\>/\>\underline{0.152}\>/\>\underline{0.122}\>/\>\underline{0.110} & 0.494 & 1 \\ \hline
\ding{51}  & \ding{55}  & \ding{55}  & 84.6\>/\>80.0\>/\>66.5\>/\>63.0\>/\>60.4 & 0.407\>/\>0.331\>/\>0.231\>/\>0.209\>/\>0.201 & 0.953 & 1 \\
\ding{51}  & \ding{55}  & \ding{51}  & - & - & - & - \\
\ding{51}  & \ding{51}  & \ding{51}  & 81.3\>/\>75.5\>/\>64.0\>/\>62.3\>/\>60.0 & 0.545\>/\>0.467\>/\>0.343\>/\>0.324\>/\>0.312 & 0.000 & 1 \\
\ding{51}  & \ding{51}  & \ding{55}  & \underline{77.0}\>/\>\underline{72.3}\>/\>\underline{56.0}\>/\>\underline{52.1}\>/\>\underline{47.8} & \textbf{0.313}\>/\>\textbf{0.221}\>/\>\textbf{0.142}\>/\>\textbf{0.113}\>/\>\textbf{0.101} & 0.586 & 1 \\ \hline
\end{tabular}%
}
\end{table}

% 为检验我们所提出的混合数据训练、无条件预训练、双分支噪声预测模型权重不冻结的有效性，我们进行了消融实验对比这三个技术方法的有效性。
% To validate the effectiveness of our method involving mixed data training, unconditional pre-training, and unfreezing of partial weights in the DBControl model, we conduct ablation experiments to compare the effectiveness of these three techniques.

We conducted ablation experiments to validate our method's effectiveness in mixed data training, unconditional pre-training, and partial weight freezing.

\textbf{Mixed data training} refers to the practice of combining a small conditioned dataset and a large unconditioned one and jointly training them. Table \ref{tab:ablation} demonstrates that this effectively reduces the FCD and MMD between the generated and target molecules, and also improves the validity of molecules without post-correction.

% 混合训练是指我们将数据量较小的有条件数据集和数据量较大的无条件数据集混合，联合训练条件生成模型。表1展示了该方法不仅可以有效提升生成的分子与目标分子之间的FCD和MMD较小，还提升了在不加入事后修正时分子的有效性。

\textbf{Unconditional pre-training} refers to training an unconditional generation model on datasets such as QM9 or ZINC250k, while simultaneously training a conditional model. While maintaining constant conditions, utilizing the unconditional model trained on the QM9 significantly improves the molecule generation quality. However, it affects the molecule validity during the initial generation process.

\textbf{Weight freezing} refers to the proposed DBControl model, which consists of two branches (i.e., $B_1$, $B_2$). First, $B_1$ undergoes unconditional training and is fine-tuned with $B_2$ during mixed conditional training. During mixed conditional training, we set the $B_1$ weights to remain unfrozen, distinguishing our method from other methods \citep{controlnet}. Various datasets play different roles, and simply freezing the weights of the pre-trained branch $B_1$ may hinder effective data distribution learning. Furthermore, drug molecule features differ from those of images, and while molecular features may be similar, their properties may not necessarily be. Therefore, weight freezing obstructs the channel for feature distribution transfer between large and small datasets designed for specific tasks.

\section{Conclusion}

This paper proposed a regressor-free guidance molecule generation model to ensure sampling within a more effective space, supporting the DRP task. Regressor-free guidance combines the score estimate of the DBControl model with the gradient of a regression controller one based on number labels. The regression controller model converts the target $\mathrm{IC}_{50}$ and cell line into text constrained by the CN-KG, effectively mapping the response value between drugs and cell lines. Additionally, to enhance noise prediction performance, we introduced the DBControl model for score estimation. The experimental results on real-world datasets during the DRP task demonstrated our method's effectiveness in de novo drug discovery, providing a novel and efficient solution for drug discovery.

\newpage

\bibliographystyle{unsrt}
\bibliography{neurips_2024}

\newpage
\appendix

% \textbf{Organization} OrganizationOrganizationOrganizationOrganizationOrganizationOrganizationOrganizationOrganizationOrganizationOrganization.

\textbf{Organization} The appendix is structured as follows: In Section \ref{A}, the authors present the proposed framework for regressor-free guidance molecule generation, delineating its three training stages. Section \ref{B} delves into related work, emphasizing the significance and practical implications of regressor-guidance molecule generation. Section \ref{D} offers proofs for the three propositions formulated in this paper. Experimental details about the generation tasks are expounded upon in Section \ref{D}, alongside supplementary experimental results. Finally, Section \ref{Visualization} provides visual representations of the generated molecules and graphs.

% The appendix is organized as follows: In Section \ref{A}, the authors introduce the proposed framework for regressor-free guidance molecule generation, outlining its three training stages. Section \ref{B} 介绍了相关工作，重点说明了回归器引导分子生成的重要意义和应用价值. Section \ref{D} 提供了本文所设置的三个命题的证明。Experimental details for the generation tasks are provided in Section \ref{D}, with additional experimental results presented. Finally, Section \ref{E} visualizes the generated molecules and graphs.

\section{Model details}
\label{A}

\subsection{Framework of regressor-free guidance molecule generation}

We propose the regressor-free guidance molecule generation, which involves three training stages.

The first stage is the training of the unconditional molecule generation model. Large-scale unconditional datasets such as QM9 \citep{QM9}, GDB-17 \citep{GDB17}, and ZINC250k \citep{ZINC250k} are used for training. During this phase, the model, specifically the noise prediction model, is trained without incorporating any specific condition information, focusing solely on learning the fundamental patterns and structures of molecule generation.

The second stage is the training of the regression controller model. Specific task datasets, such as GDSCv2 for the drug–cell line response task, are used. In this phase, contrastive learning methods are employed to train the regression controller model, converting the regression labels of drug information into text. A drug encoder and a text encoder work together to optimize this conversion process, ensuring that the response label between the drugs and the cell lines are effectively represented.

The third stage is the training and sampling with the conditional noise prediction model. This phase uses a mixture of conditional and unconditional datasets to train the noise prediction model, ensuring it can generate valid molecules under specific conditions. During the sampling process, the text conditions generated by the regression controller model guide the molecule generation process. This method employs a regressor-free guidance generation, similar to classifier-free diffusion guidance \citep{Classifier-Free}.

This framework integrates the training of the unconditional molecule generation model, the training of the regression controller model, and the training and sampling of the conditional noise prediction model. This method can sample within a smaller range near specific response values, thereby improving generation efficiency while also increasing the success rate of drug discovery.

\subsection{Schematic of DBControl model}

\begin{figure}[!h]
    \centering
    \includegraphics[width=0.8\linewidth]{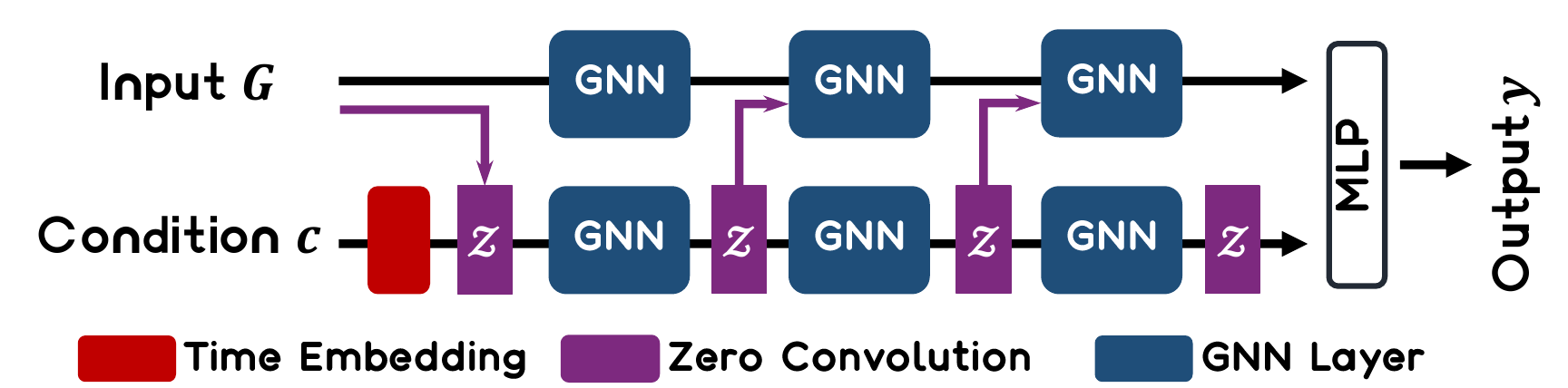}
    \caption{Schematic of the DBControl.}
    \label{fig:DControl-NP}
\end{figure}

The DBControl model, as shown in Figure \ref{fig:DControl-NP}, consists of three main components: the input layer $\mathbf{G}$ representing the molecular graph, the condition vector $\mathbf{c}$ specifying the desired properties or constraints, and the output layer $\mathbf{y}$ generating the molecule. The model operates in three stages: 

\begin{enumerate}
    \item \textbf{Graph encoder}: The input graph $\mathbf{G}$ is encoded into a fixed-dimensional representation using a graph encoder.
    \item \textbf{Condition processing}: The condition vector $\mathbf{c}$ is processed through a neural network to extract relevant features and ensure compatibility with the graph representation.
    \item \textbf{Zero convolution}: Through zero convolution, the input graph $\mathbf{G}$ is first encoded, and the condition $\mathbf{c}$ is fused with $\mathbf{G}$.
\end{enumerate}

This schematic illustrates the architecture and flow of information within the DBControl model. In our method, we create two DBControl models to predict noise for both $\mathbf{X}$ and $\mathbf{A}$.  It is worth noting that when the noise prediction for $\mathbf{X}$, the embedding of time $t$ is mixed with condition $\mathbf{c}$ to make the model sensitive to different time embeddings of condition $\mathbf{c}$ during different time steps $t$.

\section{Related work}
\label{B}

Recently, score-based generative models have been applied to image inpainting \citep{songyang}, super-resolution \citep{super-resolution}, and image translation \citep{image-translation}. However, applying these methods to the conditional generation of molecules poses several challenges. Firstly, due to the complex dependencies between nodes and edges \citep{MolDiff}, it is necessary to simultaneously consider the validity of both types of information and the relationships between them \citep{GDSS}. Additionally, when generating molecules under specific conditions, it is essential to consider the molecular property information \citep{mol_property_1}. The prevalence of activity cliffs \citep{ac} indicates that generating molecules similar to the target molecule does not guarantee similar properties.

On this basis, generalized property-based molecule generation methods can be categorized into two types. One type solely describes molecular properties (such as chemical, physical, and biological properties) \citep{mol_property_2, mol_property_3}, while the other type describes interactions, connections, or associations between molecules (including interaction forces \citep{interaction_forces}, reactions with cell lines \citep{drp}, and protein-ligand binding patterns \citep{dta}). Existing diffusion-based molecule generation methods essentially guide generation through classification \citep{DiGress}, yet these properties naturally exist in continuous numerical (regression) form.

The generation models based on classification guidance and regression guidance exhibit fundamental differences in molecule generation. Models guided by classification primarily employ the classifier to direct molecule generation, categorizing molecules into different classes or sets. This method is typically suitable for generating samples for classification tasks \citep{cls_guide_1}, such as distinguishing between drug and non-drug molecules. In contrast, regressor-based guidance method aim to generate molecules with specific numerical properties or response values. This method focuses more on the numerical features of molecules, such as predicting drug activity or optimizing specific molecular properties. Therefore, the choice between these guidance types depends on the specific application requirements and generation goals. In the majority of drug screening tasks \citep{ds_1, ds_2}, priority is given to molecules with superior properties determined through the numerical ranking of molecular attributes. This prioritization favors molecules exhibiting better properties, as assessed through virtual calculations \citep{ds_3}. Consequently, regressor-free guidance generation models are crucial in a series of tasks for drug screening.

\section{Proof of propositions}
\label{C}

\begin{propositionn}[Main proposition]
    \label{proof_lemma1}
    For any $\bm{C}_{\mathrm{aim} } \in\left ( 0,1 \right ) $,  then $\left \| S_{cls} \right \| \ge \left \| S_{reg} \right \|$  exists.
\end{propositionn}

\noindent \textit{Proof.} $\xi$ is the number of valid digits, and $\xi \in \mathbb{N} $. The significant figures are preserved by rounding off, hence the sampling radius of reg at $\bm{C}_{\mathrm{aim} }$ point is $\frac{10^{-\xi}}{2}$. Then, we have:

% $\left \| S_{cls} \right \| = (T\pm \varepsilon_1)-(0\pm \varepsilon_1)=T$ and $\left \| S_{reg} \right \|=(\bm{C}_{\mathrm{aim} }+  \frac{10^{-\xi}}{2}\pm \varepsilon _2 )- (\bm{C}_{\mathrm{aim} }-  \frac{10^{-\xi}}{2}\pm \varepsilon _2 )=10^{-\xi}$. 

\begin{align*}
\left \| S_{cls} \right \| & = (T\pm \varepsilon_1)-(0\pm \varepsilon_1) \\
& = T\\
\left \| S_{reg} \right \| & = (\bm{C}_{\mathrm{aim} }+  \frac{10^{-\xi}}{2}\pm \varepsilon _2 )- (\bm{C}_{\mathrm{aim} }-  \frac{10^{-\xi}}{2}\pm \varepsilon _2 ) \\
& = 10^{-\xi} \\
\left \| S_{cls} \right \|-\left \| S_{reg} \right \| & = T-10^{-\xi} \\
& \ge  0.
\end{align*}

\begin{propositionn}[Uniqueness of $\bm{C}_{\mathrm{aim} }$ Representation]
    \label{proof_lemma2}
    For any $\lambda \in \left [ 0,1 \right ] $,  and  $\lambda \ne \bm{C}_{\mathrm{aim} }$, then we say that  
   $\Theta(\bm{C}_{\mathrm{aim} })\ne \Theta(\lambda )$.
\end{propositionn}

\noindent \textit{Proof.} We will prove Lemma \ref{lemma:unique} by contradiction. Suppose there exists a $\lambda \in \left [ 0,1 \right ] $,  $\lambda \ne \bm{C}_{\mathrm{aim} }$, and then  $\Theta(\bm{C}_{\mathrm{aim} })  = \Theta(\lambda )$. Then, we utilize Proposition \ref{thm:Equal_Interval} to derive a variant of the left-hand side:

\begin{align*}

\Theta(\bm{C}_{\mathrm{aim}}) &= \Theta(\bm{C}_{\mathrm{aim}}+1) + \mathbf{l} +\varepsilon_3 \\
 &= \Theta(\bm{C}_{\mathrm{aim}}+2) + 2\mathbf{l} + 2\varepsilon_3 \\
 & \vdots \\
 &= \Theta(\lambda) + \left|\lambda - \bm{C}_{\mathrm{aim}}\right| \mathbf{l} + \left|\lambda - \bm{C}_{\mathrm{aim}}\right|\varepsilon_3,
\end{align*}

Then, we have:

\begin{equation}
    \label{wrong_eq}
    \Theta(\lambda) = \Theta(\lambda) + \left|\lambda - \bm{C}_{\mathrm{aim}}\right| \mathbf{l} + \left|\lambda - \bm{C}_{\mathrm{aim}}\right|\varepsilon_3.
\end{equation}

The equation \ref{wrong_eq} holds under two conditions.  First, $\lambda = \bm{C}_{\mathrm{aim}}$, which contradicts the assumption of this proposition. Second, $\mathbf{l}+\varepsilon_3
= \mathbf{0}$. Since $\varepsilon_3$ is much smaller than $\mathbf{l}$, and $\left| \mathbf{l} \right|=1$, this solution is not valid. Considering both solutions contradict the known conditions, the negative proposition of Proposition \ref{lemma:unique} is false, thus Proposition \ref{lemma:unique} holds.

\begin{propositionn}[Equal interval representation of $\Theta$]
    \label{proof_lemma3}
   For any $\xi$,  a perturbation $\varepsilon_3 $ exist to make $ \Theta(V[i])-\Theta(V[i+1])  = \mathbf{l} + \varepsilon_3$.
\end{propositionn}

\noindent \textit{Proof.} To consider $\mathcal{L}_{\mathrm{KGE}}$ with the squared euclidean distance as dissimilarity function \citep{TransE}, we have:

\begin{equation}
\mathcal{L}_{\mathrm{KGE}}= \sum_{ (h,l,t)\in S   } \left [ \gamma +d(\textbf{h}+\textbf{l},\textbf{t}) -d(\textbf{t}+\textbf{l} ,\textbf{h})\right ]_+.
\end{equation}

% \begin{equation}
% d(\mathbf{h} +\mathbf{l} ,\mathbf{t} )=\left \| \mathbf{h} \right \| ^2_2+\left \| \mathbf{l} \right \| ^2_2+\left \| \mathbf{t}\right \| ^2_2 -2(\mathbf{h}^T \mathbf{t} +\mathbf{l}^T(\mathbf{t}- \mathbf{h}))=\varepsilon_3
% \end{equation}

Then, $ \Theta(V[i+1])$ and $\Theta(V[i])$ can be represent as $\mathbf{h}$ and $\mathbf{t}$. As the translation-based model proposed by \citep{TransE}, $\mathbf{t}$ should be a nearest neighbor of $\mathbf{h}+\mathbf{l}$, i.e. $\mathbf{h}+\mathbf{l}\approx  \mathbf{t} $. Then, the $ \Theta(V[i])+\mathbf{l} \approx \Theta(V[i+1])$. By utilizing $\varepsilon_3$ to denote the error term of $\mathcal{L}_{\mathrm{KGE}}$, we have $ \Theta(V[i+1])+\mathbf{l} +\varepsilon_3 = \Theta(V[i])$.

\section{Experimental details}
\label{D}
\subsection{Baselines}
\label{A_Baselines}
We selected representative diffusion methods with high performance for molecule generation in recent years, including GDSS \citep{GDSS}, EDM \citep{E3Diffusion}, CDGS \citep{CDGS}, GeoLDM \citep{GeoLDM}, DiGress \citep{DiGress}, MOOD \citep{MOOD}, GruM-2D \citep{GruM2D}.

\subsection{Datasets}
The dataset information, summarized in Table \ref{tab:mol_dataset_uncondition} and Table \ref{tab:mol_dataset_condition}, comprises openly accessible datasets available online for academic purposes. All datasets listed are freely downloadable and intended for academic use.

\begin{table}[!htbp]
\centering
\setlength{\tabcolsep}{5pt}
\caption{Unconditional Molecule dataset information.}
\label{tab:mol_dataset_uncondition}
\renewcommand\arraystretch{1.4}
\resizebox{\columnwidth}{!}{%
\begin{tabular}{ccccc}
\hline
Dataset  & Number of molecules & Number of nodes  & Number of node types & Number of edge types \\ \hline
ZINC250k \citep{ZINC250k} & 249,455             & $6\leq|V|\leq38$ & 9                    & 3                  \\
QM9 \citep{QM9}     & 133,885             & $1\leq|V|\leq9$  & 4                    & 3                   \\ 
GDB-17  \citep{GDB17}    & 166,443,860,262             & -  & 17                    & -                   \\ \hline
\end{tabular}%
}
\end{table}

\begin{table}[!htbp]
    \centering
    \renewcommand\arraystretch{1.5}
    \setlength{\tabcolsep}{12pt}
    \caption{Datasets for DRP tasks, where sensitivity assay means the method for calculating the $\mathrm{IC}_{50}$.}
    \resizebox{\textwidth}{!}{%
    \begin{tabular}{cccccc}
        \hline
        Dataset     &
        \multicolumn{1}{c}{\begin{tabular}[c]{@{}c@{}}GDSCv1\\ \citep{GDSCv1}\end{tabular}}
        &
        \multicolumn{1}{c}{\begin{tabular}[c]{@{}c@{}}GDSCv2\\ \citep{GDSC}\end{tabular}}
        &
        \multicolumn{1}{c}{\begin{tabular}[c]{@{}c@{}}CTRPv1\\ \citep{CTRPv1}\end{tabular}}
        &
        \multicolumn{1}{c}{\begin{tabular}[c]{@{}c@{}}CTRPv2\\ \citep{CTRPv2_1}\end{tabular}}
        &
        \multicolumn{1}{c}{\begin{tabular}[c]{@{}c@{}}GCSI\\ \citep{GCSI_1}\end{tabular}}
        \\  \hline
        Total Number     & 258196   & 152839   & 30545  & 130855   & 6178      \\ 
        Drug Type Number & 309      & 223      & 354    & 545      & 43        \\ 
        Average per drug & 835      & 655      & 86    & 240      & 143           \\
        Cell Type Number & 968      & 990      & 184    & 589      & 366           \\
        Sensitivity assay & Syto60      & CellTitreGlo      & CellTitreGlo    & CellTitreGlo      & CellTitreGlo           \\
        Model Evaluation & Hold-out & Hold-out & K-Flod & Hold-out & K-Flod \\ \hline
    \end{tabular}%
    }
    \label{tab:mol_dataset_condition}
\end{table}

\subsection{Settings}
\label{appendix:settings}
We use NVIDIA RTX A6000 for model training, requiring support for 48GB of VRAM and a memory capacity of no less than 128 GB. The maximum number of atoms accepted as input molecules and also the maximum number of atoms in generated molecules is denoted as $\mathrm{MNN}$, $\mathrm{MFN}$ is the maximum number of atom types.

\textbf{The regression controller model.} The margin for $\mathcal{L}_{\mathrm{KGE} }$ is set to 1.0, the entity dimension (dim) is set to 128, and the learning rate is set to  0.0001.  For $\mathcal{L}_{\mathrm{KGE} }$, the batch size is 1024, the $\mathrm{MNN}$ is 100, and the $\mathrm{MFN}$ is 10. The Adam optimizer is used with a weight decay of 0.05.

\textbf{The dual-branch controlled noise prediction model.} In the diffusion phase, the DBControl model undergoes mixed training with a learning rate of 0.0001 and a weight decay of 0.0005 using the Adam optimizer. The batch size is set to 512, the $\mathrm{MNN}$ set to 100, and $\mathrm{MFN}$ to 10. The model is initialized with pre-trained weights from QM9, with the pre-training parameters being identical to those used in the conditional mixed training. In the sampling phase, the signal-to-noise ratio (SNR) is set to 0.2, and the scale epsilon is 0.7. The total number of time steps is 1000.

\textbf{Compute resource requirements and time consumption.} During diffusion, with a batch size of 512, approximately 30,000 MB of memory is required, and it takes 96 hours for 500 epochs. Sampling 1000 molecules requires about 10,000 MB of memory and 0.5 hours.

\subsection{Metrics}
\label{appendix:metrics}
We evaluate the quality of the 1,000 generated molecules with the following metrics. FCD \citep{FCD} evaluates the distance between the training and generated sets using the activations of the penultimate layer of the ChemNet. (NSPDK) MMD \citep{MMD} is the MMD between the generated molecules and test molecules which takes into account both the node and edge features for evaluation.  Specifically, FCD measures the ability in the view of molecules in chemical space, while NSPDK MMD measures the ability in the view of the graph structure \citep{GDSS}.

\subsection{Error assessment}
\label{apx:ErrorAssessment}

We conducted 5 random seed sampling experiments for each of the 4 cell lines involved in the overall experiment in Section \ref{exp:overall} to assess the errors of our method (as shown in Tables \ref{tab:var_1}, \ref{tab:var_2}, \ref{tab:var_3}, and \ref{tab:var_4}). These tables displays the mean and variance of the sampling results for the FCD and MMD metrics over 5 random seeds. Both the mean and variance decrease as TopK increases. A smaller TopK indicates more stringent evaluation criteria; for instance, TopK=1 implies that the molecular generation method must be similar to a single target molecule and must be constrained under the conditions of the target cell line and target $\mathrm{IC}_{50}$. The consistent trend in mean and variance also reflects the alignment of our method with established facts and its effectiveness.

Our coefficient of variation, approximately 0.041\%, 0.028\%, 0.037\%, 0.036\%, 0.034\% for FCD and 0.002\%, 0.003\%, 0.002\%, 0.001\%, 0.002\% for MMD, respectively, indicates that our method produces molecules with high stability and low error.

\begin{table}[htpb]
\centering
\setlength{\tabcolsep}{15pt}
\caption{ Five random seed experiments regarding the target $\mathrm{IC}_{50}$ score at 0.4 for the cell line ES5 at $T = 1000$ time steps.
}
\label{tab:var_1}
\renewcommand\arraystretch{1.25}
\resizebox{\columnwidth}{!}{%
\begin{tabular}{ccccccccc}
\hline
\multirow{2}{*}{Metrics} & \multirow{2}{*}{TopK} & \multicolumn{5}{c}{Seed} & \multirow{2}{*}{Mean} & \multirow{2}{*}{Var ($\pm$)} \\ \cline{3-7}
 &  & 40 & 41 & 42 & 43 & 44 &  &  \\ \hline
\multirow{5}{*}{FCD} & 3 & 68.815 & 69.241 & 68.867 & 68.997 & 68.977 & 68.9794 & 0.027103 \\
 & 5 & 63.321 & 63.547 & 63.371 & 63.358 & 63.433 & 63.406 & 0.007841 \\
 & 10 & 53.549 & 53.84 & 53.653 & 53.731 & 53.966 & 53.7478 & 0.026226 \\
 & 15 & 50.716 & 50.987 & 50.711 & 50.823 & 50.998 & 50.847 & 0.019659 \\
 & 20 & 55.003 & 55.242 & 54.977 & 55.057 & 55.271 & 55.11 & 0.018823 \\ \hline
\multirow{5}{*}{MMD} & 3 & 0.33 & 0.324 & 0.327 & 0.329 & 0.326 & 0.3272 & 5.7E-06 \\
 & 5 & 0.228 & 0.226 & 0.226 & 0.228 & 0.226 & 0.2268 & 1.2E-06 \\
 & 10 & 0.136 & 0.137 & 0.135 & 0.136 & 0.136 & 0.136 & 5E-07 \\
 & 15 & 0.117 & 0.117 & 0.116 & 0.116 & 0.117 & 0.1166 & 3E-07 \\
 & 20 & 0.107 & 0.108 & 0.107 & 0.106 & 0.107 & 0.107 & 5E-07 \\ \hline
\end{tabular}%
}
\end{table}

\begin{table}[htpb]
\centering
\caption{ Five random seed experiments regarding the target $\mathrm{IC}_{50}$ score at 0.4 for the cell line Hs-578-T at $T = 1000$ time steps.
}
\label{tab:var_2}
\setlength{\tabcolsep}{15pt}
\renewcommand\arraystretch{1.25}
\resizebox{\columnwidth}{!}{%
\begin{tabular}{ccccccccc}
\hline
\multirow{2}{*}{Metrics} & \multirow{2}{*}{TopK} & \multicolumn{5}{c}{Seed} & \multirow{2}{*}{Mean} & \multirow{2}{*}{Var ($\pm$)} \\ \cline{3-7}
 &  & 40 & 41 & 42 & 43 & 44 &  &  \\ \hline
\multirow{5}{*}{FCD} & 3 & 98.5 & 98.16 & 98.384 & 98.241 & 98.039 & 98.2648 & 0.033016 \\
 & 5 & 68.985 & 68.784 & 68.811 & 69.016 & 68.634 & 68.8460 & 0.024559 \\
 & 10 & 67.985 & 67.936 & 67.714 & 68.061 & 67.628 & 67.8648 & 0.034207 \\
 & 15 & 64.599 & 64.513 & 64.35 & 64.653 & 64.226 & 64.4682 & 0.031475 \\
 & 20 & 60.327 & 60.328 & 60.101 & 60.334 & 60.008 & 60.2196 & 0.023803 \\ \hline
\multirow{5}{*}{MMD} & 3 & 0.294 & 0.295 & 0.299 & 0.296 & 0.299 & 0.2966 & 5.3E-06 \\
 & 5 & 0.199 & 0.198 & 0.2 & 0.2 & 0.201 & 0.1996 & 1.3E-06 \\
 & 10 & 0.159 & 0.157 & 0.158 & 0.161 & 0.159 & 0.1588 & 2.2E-06 \\
 & 15 & 0.13 & 0.128 & 0.129 & 0.13 & 0.129 & 0.1292 & 7E-07 \\
 & 20 & 0.109 & 0.108 & 0.109 & 0.11 & 0.11 & 0.1092 & 7E-07 \\ \hline
\end{tabular}%
}
\end{table}

% \usepackage{multirow}
% \usepackage{graphicx}
\begin{table}[htpb]
\centering
\caption{Five random seed experiments regarding the target $\mathrm{IC}_{50}$ score at 0.4 for the cell line ES3 at $T = 100$ time steps.
}
\label{tab:var_3}
\setlength{\tabcolsep}{15pt}
\renewcommand\arraystretch{1.25}
\resizebox{\columnwidth}{!}{%
\begin{tabular}{ccccccccc}
\hline
\multirow{2}{*}{Metrics} & \multirow{2}{*}{TopK} & \multicolumn{5}{c}{Seed} & \multirow{2}{*}{Mean} & \multirow{2}{*}{Var ($\pm$)} \\ \cline{3-7}
 &  & 40 & 41 & 42 & 43 & 44 &  &  \\ \hline
\multirow{5}{*}{FCD} & 3 & 76.997 & 76.962 & 76.928 & 76.512 & 76.723 & 76.8244 & 0.041833 \\
 & 5 & 71.863 & 71.965 & 71.879 & 71.593 & 71.637 & 71.7874 & 0.026515 \\
 & 10 & 55.041 & 55.094 & 55.000 & 54.934 & 54.867 & 54.9872 & 0.007936 \\
 & 15 & 51.622 & 51.546 & 51.649 & 51.405 & 51.410 & 51.5264 & 0.01321 \\
 & 20 & 47.072 & 47.133 & 47.113 & 46.923 & 46.949 & 47.0380 & 0.009238 \\ \hline
\multirow{5}{*}{MMD} & 3 & 0.306 & 0.304 & 0.303 & 0.306 & 0.308 & 0.3054 & 3.8E-06 \\
 & 5 & 0.208 & 0.210 & 0.210 & 0.214 & 0.212 & 0.2108 & 5.2E-06 \\
 & 10 & 0.130 & 0.129 & 0.129 & 0.129 & 0.131 & 0.1296 & 8E-07 \\
 & 15 & 0.104 & 0.103 & 0.103 & 0.103 & 0.106 & 0.1038 & 1.7E-06 \\
 & 20 & 0.092 & 0.091 & 0.091 & 0.091 & 0.094 & 0.0918 & 1.7E-06 \\ \hline
\end{tabular}%
}
\end{table}

% \usepackage{multirow}
% \usepackage{graphicx}
% \begin{table}[]
% \centering
% \caption{cell	688007	NCI-H187
% ic50	0.35	T=1000

% }

\begin{table}[htpb]
\centering
\caption{Five random seed experiments regarding the target $\mathrm{IC}_{50}$ score at 0.35 for the cell line NCI-H187 at $T = 1000$ time steps.
}
\label{tab:var_4}
\setlength{\tabcolsep}{15pt}
\renewcommand\arraystretch{1.25}
\resizebox{\columnwidth}{!}{%
\begin{tabular}{ccccccccc}
\hline
\multirow{2}{*}{Metrics} & \multirow{2}{*}{TopK} & \multicolumn{5}{c}{Seed} & \multirow{2}{*}{Mean} & \multirow{2}{*}{Var ($\pm$)} \\ \cline{3-7}
 &  & 40 & 41 & 42 & 43 & 44 &  &  \\ \hline
\multirow{5}{*}{FCD} & 3 & 90.54 & 90.541 & 90.225 & 90.352 & 90.7 & 90.4716 & 0.034194 \\
 & 5 & 86.372 & 86.412 & 86.269 & 86.22 & 86.621 & 86.3788 & 0.024271 \\
 & 10 & 76.51 & 76.386 & 76.189 & 76.257 & 76.555 & 76.3794 & 0.024792 \\
 & 15 & 74.333 & 74.307 & 74.132 & 74.043 & 74.401 & 74.2432 & 0.02237 \\
 & 20 & 68.846 & 68.802 & 68.564 & 68.526 & 68.895 & 68.7266 & 0.028745 \\ \hline
\multirow{5}{*}{MMD} & 3 & 0.432 & 0.431 & 0.428 & 0.429 & 0.426 & 0.4292 & 5.7E-06 \\
 & 5 & 0.282 & 0.275 & 0.273 & 0.274 & 0.271 & 0.2750 & 1.75E-05 \\
 & 10 & 0.186 & 0.179 & 0.178 & 0.178 & 0.178 & 0.1798 & 1.22E-05 \\
 & 15 & 0.151 & 0.148 & 0.146 & 0.148 & 0.146 & 0.1478 & 4.2E-06 \\
 & 20 & 0.128 & 0.123 & 0.123 & 0.124 & 0.123 & 0.1242 & 4.7E-06 \\ \hline
\end{tabular}%
}
\end{table}

\section{Visualization}

\label{Visualization}

\begin{figure}[htbp]
  \centering
  \begin{subfigure}[b]{0.49\textwidth}
    \centering
    \includegraphics[width=\textwidth]{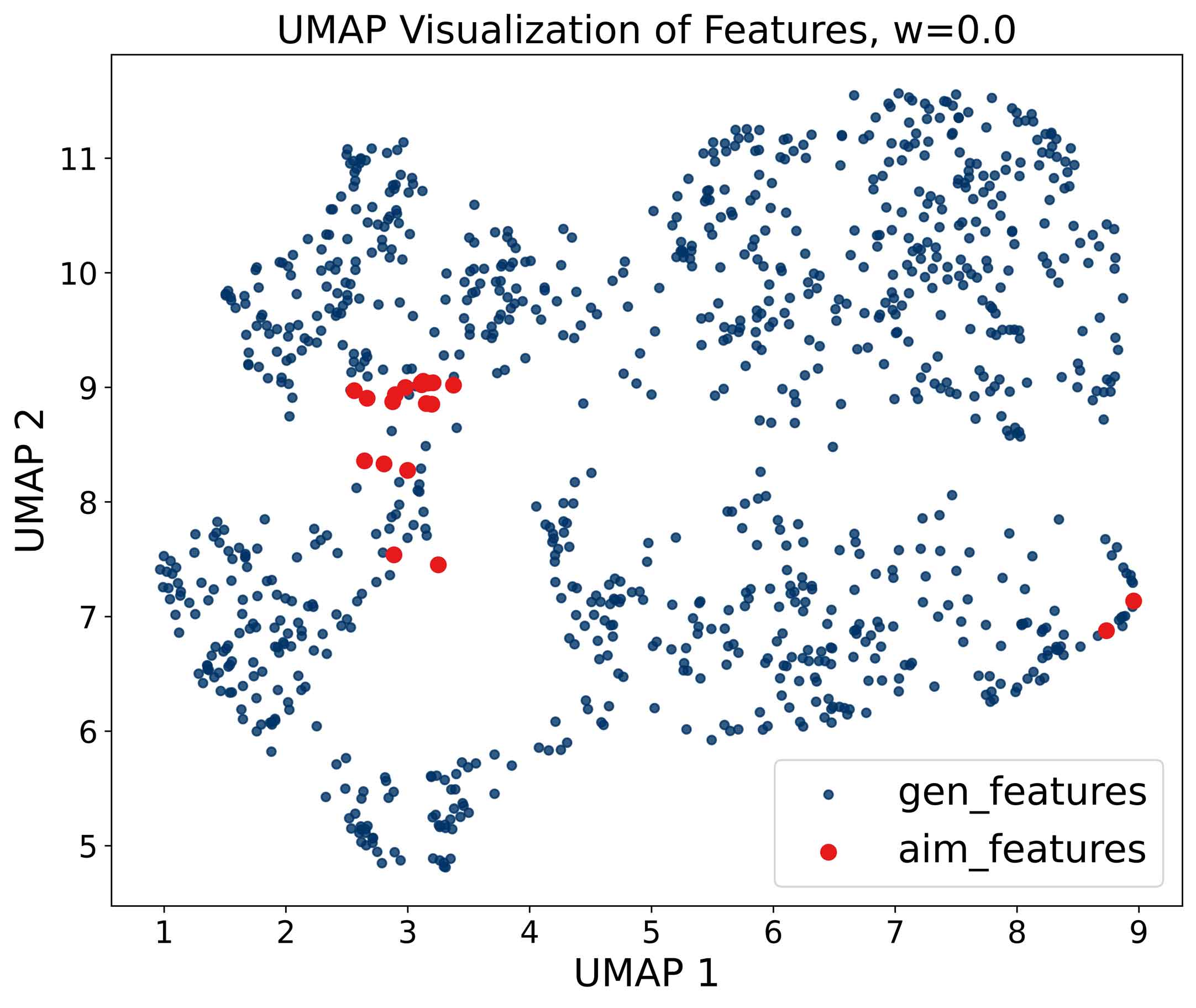}
    \caption{Conditional Generation with w is 0}
    \label{fig:umap_sub1}
  \end{subfigure}
  \hfill
  \begin{subfigure}[b]{0.49\textwidth}
    \centering
    \includegraphics[width=\textwidth]{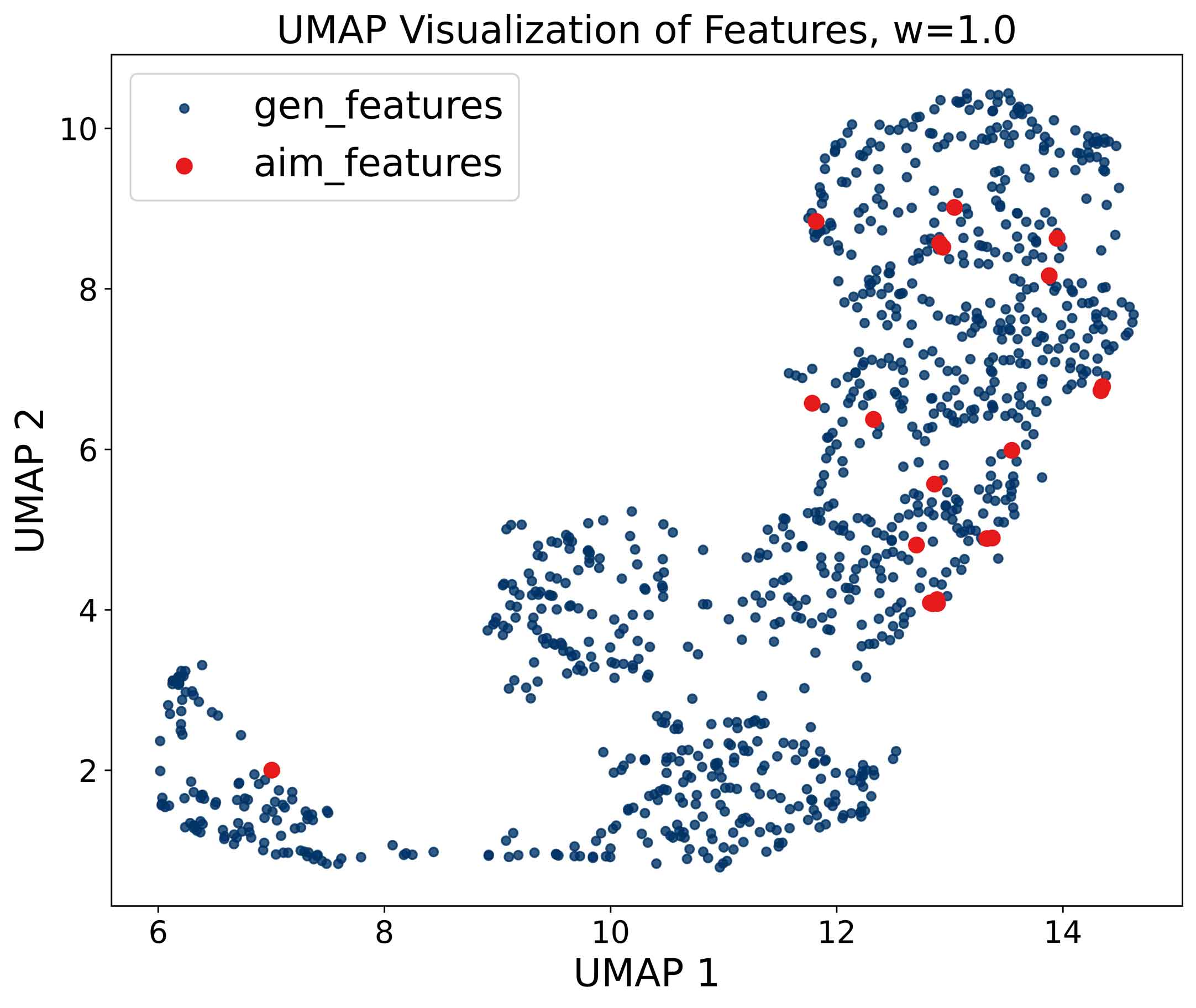}
    \caption{Conditional Generation with w is 1}
    \label{fig:umap_sub2}
  \end{subfigure}
  \bigskip
  
  \begin{subfigure}[b]{0.80\textwidth}
    \centering
    \includegraphics[width=\textwidth]{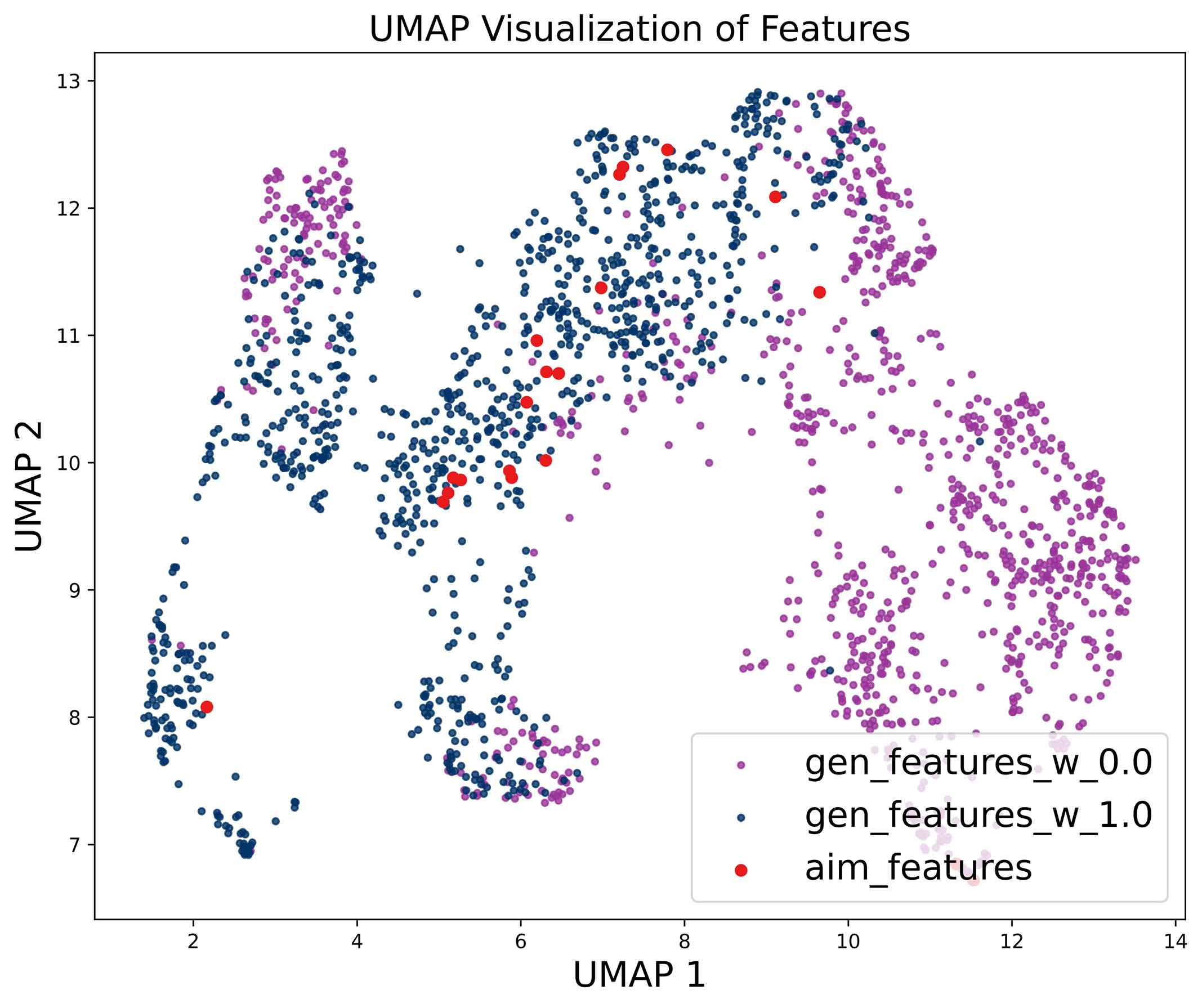}
    \caption{Comparison of molecule generation with and without conditional guidance}
    \label{fig:umap_sub3}
  \end{subfigure}
  \caption{Molecule generation: conditional vs. unconditional guidance}
  \label{fig:umap}
\end{figure}

We visualize the features of molecule sampling under different conditions to compare the sampling effects. Taking the NCI-H187 cell line (ID: 688007) as an example, the target $\mathrm{IC}_{50}$ is set to 0.35. The reason for setting it to 0.35 instead of the common 0.4 in previous cases is that this cell line has experimental data around 0.35. Fig. \ref{fig:umap}(a) and Fig. \ref{fig:umap}(b) show the feature distributions of generated molecules and target molecules under condition strengths $w = 0$ and $w = 1$, respectively. We selected the 20 target molecules nearest to the $\mathrm{IC}_{50}$ in the molecular reaction data corresponding to this cell line in the dataset. Features were extracted using FCDNet \citep{FCD} with a dimensionality of 512, followed by dimensionality reduction by Uniform Manifold Approximation and Projection (UMAP). When the condition strength $w = 1$, the feature distribution of generated molecules is more clustered around the target molecules compared to $w = 0$.

In addition, Fig. \ref{fig:umap}(c) uses a similar method to generate feature distribution plots. We simultaneously reduce the dimensionality of $w = 0$ and $w = 1$ cases to the same feature space for visualization. The condition-guided generation results are closer to the target molecules compared to the uncondition-guided generation results. Even in the lower left corner of Fig. \ref{fig:umap}(c), where the target molecules are located, the condition-guided generated molecules are also distributed.

\begin{figure}
    \centering
    \includegraphics[width=1\linewidth]{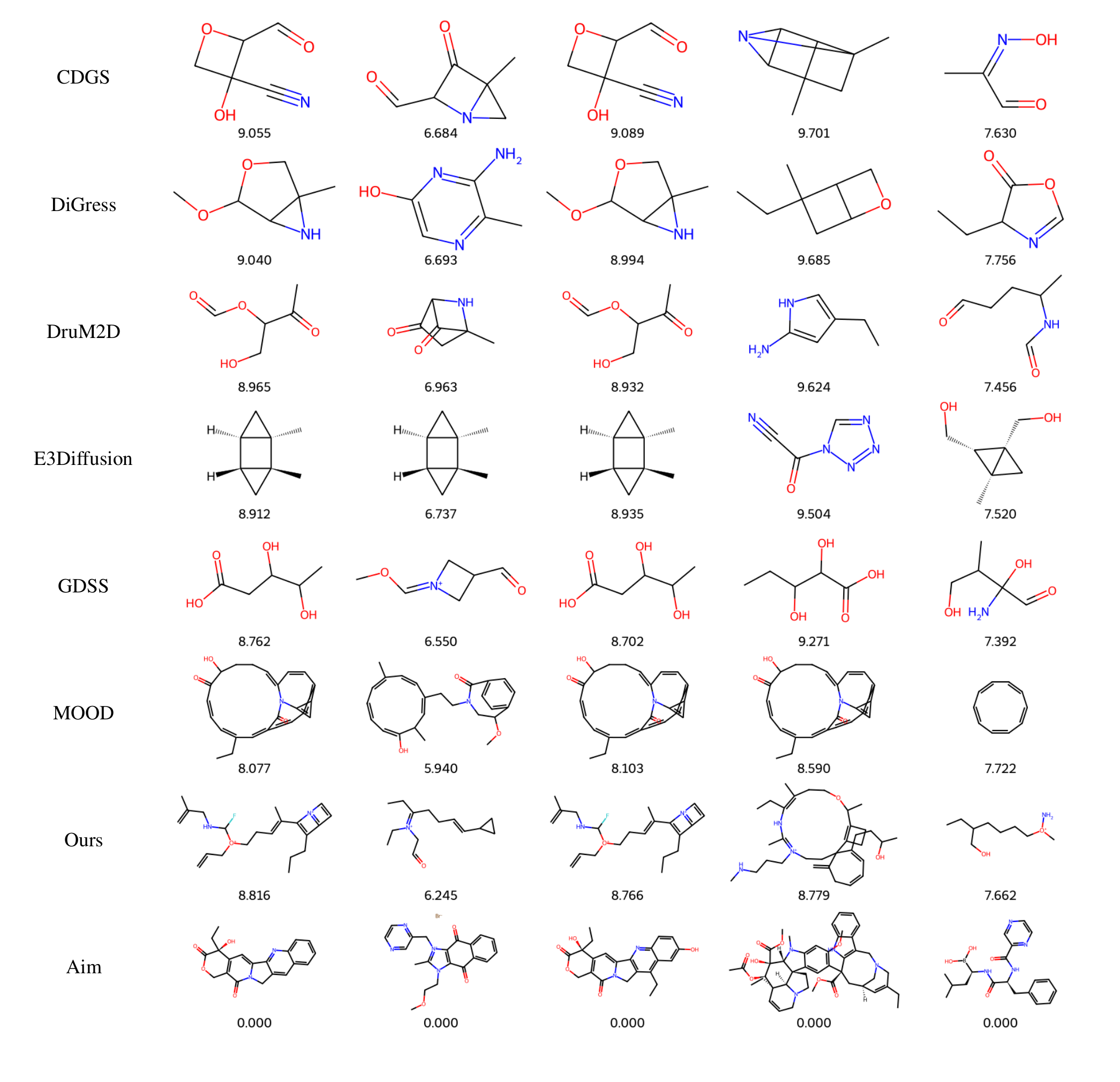}
    \caption{The comparison of molecule images for seven diffusion-based molecule generation methods. Among 1,000 generated molecules using the target pair (NCI-H187, $\mathrm{IC}_{50}$=0.35), the one most similar to the target molecule is selected.}
    \label{fig:mol_v}
\end{figure}

Fig. \ref{fig:mol_v} illustrates the structures of the five molecules most similar to the five target molecules generated by seven different methods. We observe that although MOOD \citep{MOOD} shows advantages in similarity, the generated molecules all exhibit large cyclic structures of carbon, which significantly deviate from the structure of the target molecules, making it uncertain whether they are effective molecules. Other methods generate molecules with distinct substructures; for instance, E3Diffusion \citep{E3Diffusion} generates molecules with similar scaffolds, while GDSS \citep{GDSS} generates molecules consisting mainly of single carbon chains. Evaluating molecular generation performance solely based on metrics is not reliable, and visualizing the generated results cannot serve as an objective measure. However, based on the overall structure of the molecules, our method demonstrates strong competitiveness, whether assessed by metrics or visualized results.

\section{Future work and limitations}

\subsection{Limitations} 
\label{Appendix:Limitations}
While our work introduces a novel method and demonstrates significant improvements in the field of conditional molecular graph generation for DRP tasks, there are several limitations to acknowledge:

\textbf{Lack of Wet Lab Validation.} Our method has not yet been validated through wet lab experiments. This limits the practical verification of the generated molecules' efficacy and safety in real-world biological settings. However, conducting wet lab experiments poses significant challenges for us due to the unique nature of the molecular generation field. Wet lab experiments require teams to have access to facilities for molecular synthesis, cell culture, and biological experimentation. These experiments are costly and have extremely long trial periods. Even the largest drug discovery company globally finds it difficult to afford and turns to computer-aided drug discovery instead.

\textbf{Evaluation Metrics.} While we demonstrate improved performance using standard evaluation metrics (FCD and MMD, see Appendix \ref{appendix:metrics}), these metrics may not capture all aspects of model performance, particularly in real-world applications. Additional metrics and qualitative assessments could provide a more comprehensive evaluation of the model's utility and effectiveness.

\textbf{Computational Resources.} The proposed method, including regressor-free guidance and the DBControl model, involves complex computations that demand significant computational resources, potentially limiting its feasibility in resource-constrained environments.

\subsection{Future work} 

Our proposed regressor-free guidance molecule generation will significantly contribute to accelerating the pace of drug discovery to a considerable extent. This is because we fully consider the challenges of drug discovery in molecular preliminary screening. Existing methods, which are limited to unconditional generation or classification-guided generation, fall short of practical application. Real-world applications require models with numerical, ranking, and recommendation capabilities. Therefore, we call for more scholars to explore regression tasks. In our future work, we will incorporate additional conditions into this method and validate the effectiveness of the method through wet experiments.

%%%%%%%%%%%%%%%%%%%%%%%%%%%%%%%%%%%%%%%%%%%%%%%%%%%%%%%%%%%%

% \section*{References}

%%%%%%%%%%%%%%%%%%%%%%%%%%%%%%%%%%%%%%%%%%%%%%%%%%%%%%%%%%%%

% \appendix

% \section{Appendix / supplemental material}

% Optionally include supplemental material (complete proofs, additional experiments and plots) in appendix.
% All such materials \textbf{SHOULD be included in the main submission.}

% %%%%%%%%%%%%%%%%%%%%%%%%%%%%%%%%%%%%%%%%%%%%%%%%%%%%%%%%%%%%

\end{document}